\pdfoutput=1
\documentclass[prd,aps,nofootinbib]{revtex4-2}

\usepackage[colorlinks=true,citecolor=blue,urlcolor=blue,linkcolor=blue,pdfstartview=FitH,bookmarksopen]{hyperref}
\usepackage{amsmath,amssymb}
\usepackage{bm}
\usepackage{comment}
\renewcommand\d{\partial}
\newcommand\+\dagger
\newcommand\<{\langle}
\renewcommand\>{\rangle}
\renewcommand\Im{\mathop{\mathrm{Im}}}
\newcommand\x{{\vec x}}
\newcommand\y{{\vec y}}
\newcommand\z{{\vec z}}
\newcommand\p{{\vec p}}
\newcommand\q{{\vec q}}
\newcommand\xp{{\vec x}^{\,\prime}}
\newcommand\yp{{\vec y}^{\,\prime}}

\begin{document}
\title{Applied nonrelativistic conformal field theory:\\ scattering-length and effective-range corrections to unnuclear physics}

\author{Subham Dutta Chowdhury, Ruchira Mishra, and Dam Thanh Son}
\affiliation{Kadanoff Center for Theoretical Physics, University of Chicago, Chicago, Illinois 60637, USA}

\date{September 2023}

\begin{abstract}
Due to an accidentally large $s$-wave scattering length, in a relatively wide range of energy, neutrons are approximately described by the nonrelativistic conformal field theory of unitarity fermions, perturbed by one relevant and an infinite number of irrelevant operators.  We develop a formalism which provides a nonperturbative definition of local operators in that nonrelativistic conformal field theory.  We compute the scattering-length and effective-range corrections to the two-point functions of primary charge-three operators using the technique of conformal perturbation theory.  These calculations allow us to find the first corrections to the scale-invariant behavior of the rate of nuclear reactions with three neutrons in the final state in the regime when the neutrons have small relative momenta.  
\end{abstract}

\maketitle

\tableofcontents
\section{Introduction}

It is well known that neutrons present a natural realization of fermions near unitarity: the low-energy neutron-neutron scattering is resonant in the $s$-wave and characterized by an anomalously large scattering length, $a\approx-19$~fm, as compared to a more typical (for nuclear interactions) value for the effective range $r_0\approx2.75$~fm~\cite{Slaus:1989tsq}. Fermions at unitarity is described by a universal interacting nonrelativistic conformal field theory (NRCFT)~\cite{Mehen:1999nd,Nishida_2007}, where the symmetries impose nontrivial constraints on the correlation functions.  In particular, the correlation functions possess scale invariance. The scale-invariant behavior of the neutron-neutron interaction is, however, limited to a finite energy window between $\epsilon_a\equiv\hbar^2/(m_n a^2)\sim 0.1$~MeV and $\epsilon_{r_0}\equiv\hbar^2/(m_n r_0^2)\sim 5$~MeV (where $m_n$ is the neutron mass); the upper end of this energy range is somewhat below the energies typically encountered in nuclear physics.  One system where neutrons can be approximated as fermions near unitarity is weakly-bound nuclei, especially, weakly-bound two-neutron halo nuclei. For the latter, quantitative predictions for the properties of the system can be reliably made~\cite{Hongo:2022sdr}.

It has recently been pointed out that the scale-invariant regime in neutron physics also exhibits itself in the physics of nuclear reactions~\cite{Hammer_2021}. This can be illustrated in the example of a reaction in which the final states contain several neutrons and one another particle:
\begin{equation}
  A_1 + A_2 \to B + \underbrace{n+n+\cdots}_{N~\text{neutrons}} .
\end{equation}
(An example of such a reaction, considered in Ref.~\cite{Hammer_2021}, is the radiative capture of a pion by a triton nucleus: $\pi^-+{^3}\text{H}\to \gamma+ 3n$.) At a given center-of-mass energy of the colliding particles, the spectrum of $B$ has an end point (i.e., a maximal energy) $E_0$, and near this end point the neutrons move along almost the same direction with approximately the same velocity. In this regime the reaction can be thought of as occuring in two steps. In the first, primary, step, the initial particles $A_1$ and $A_2$ collide to produce the particle $B$ and an object called ``unnucleus" $\mathcal U$, which has the same quantum numbers (mass, charge, and baryon number) as $N$ neutrons.  The unnucleus is not, however, a particle, but is a local field in the nonrelativistic conformal field theory. In the second step, the unnucleus decays into $N$ free neutrons.  
Near the end point the differential cross section is proportional to the decay rate of the unnucleus $\mathcal U$ into neutrons, which is determined by scale invariance.  The differential cross section thus vanishes according to a power law where the exponent is the scaling dimension of an operator in the nonrelativistic conformal field theory of unitarity fermions,
\begin{equation}\label{unnuclear}
  \frac{d\sigma}{d E} \sim (E_0-E)^{\Delta_{\mathcal U}-5/2}.
\end{equation}
Here $\Delta_{\mathcal U}$ is the scaling dimension of the operator in the NRCFT that produces that unnucleus $\mathcal U$ that eventually decays into $N$ final-state neutrons. The value of $\Delta_N$ for a few small values of $N$ are: $\Delta_2=2$, $\Delta_3\approx 4.27272$ in $p$-wave and 4.66622 in $s$-wave, and $\Delta_4\approx 5.0$.  This ``unnuclear'' behavior of the cross section is a nonrelativistic version of the ``unparticle physics'' speculated to exist beyond the Standard Model \cite{Georgi:2007ek}. 

One expects that the power-law ``unnuclear'' behavior derived in Ref.~\cite{Hammer_2021} has a finite range of applicability: the ``relative'' kinetic energy of the neutrons, i.e., the total kinetic energy of the neutrons in their common center-of-mass frame, has to be between $\epsilon_a$ and $\epsilon_{r_0}$.  In this paper, we compute the deviation from the scaling behavior~(\ref{unnuclear}) that comes from two sources of violation of scale invariance: the finite scattering length and the nonzero effective range in the $s$-wave neutron-neutron scattering. 

The method used in this paper is a nonrelativistic version of conformal perturbation theory.  In this approach, one uses the NRCFT of fermions at unitarity as an unperturbed starting point, and then treats violation of scale invariance as small perturbations, over which one develops a perturbation theory.  The first correction to the two-point function $\<\mathcal U\mathcal U^\+\>$ is then expressed in term of a three-point function $\<\mathcal U \mathcal L_1 \mathcal U^\+\>$, where $\mathcal L_1$ is the perturbation of the Lagrangian. 

The result can be summarized as follows.  When the relative energy of the neutrons (i.e., the energy of the neutrons in their center-of-mass frame) $\omega$ approaches the lower end of the conformal window, the correction from the scattering length becomes more and more important.  The first correction is proportional to the inverse scattering length and has the form
\begin{equation}\label{ImG-a}
  \Im G_{\mathcal U}(\omega) = 
 c_0 \omega^{\Delta_{\mathcal U-5/2}}
  \left( 1+  \frac{C_{\mathcal U}}{a\sqrt{m_n\omega}}\right) ,
\end{equation}
where $c_0$ is a normalization factor that can be absorbed in the definition of the operator $\mathcal U$.  We will compute the coefficient $C_{\mathcal U}$ for the two lowest-dimension charge-three operators\footnote{We use ``charge'' and ``particle number'' interchangeably.}: the $p$-wave ($\Delta_\mathcal{U}=4.27272$) and $s$-wave ($\Delta_\mathcal{U}=4.66622$) operators.  Numerically,
\begin{equation}
  C_{\mathcal U} = \begin{cases} 2.64179, & l = 1, \\
  4.14759, & l = 0.\end{cases}  
\end{equation}
Note that for neutrons $a<0$, so the correction in Eq.~(\ref{ImG-a}) suppresses the result, and the suppression is stronger at lower $\omega$.  This matches with the expectation that in the limit $\omega a^2\ll1$ the behavior of $\Im G_{\mathcal U}$ is dictated by the scaling dimensions in free field theory ($\Delta_{\mathcal U}=11/2$ for the $p$-wave and $13/2$ for the $s$-wave), which are larger than the corresponding values at the interacting fixed point .

When $\omega$ approaches the upper end of the conformal window $\epsilon_{r_0}$, one expects, from dimensional analysis, a correction of the form
\begin{equation}
  \Im G_{\mathcal U}(\omega) = 
 c_0 \omega^{\Delta_{\mathcal U}-5/2}
  \bigl( 1+  C'_{\mathcal U}r_0\sqrt{m_n\omega}\bigr) .
\end{equation}
Our calculation yields $C_{\mathcal U}'=0$, which means that the first effective-range correction appears at order $r_0^2 m_n \omega$.  We do not currently completely understand the reason for the absence of the $O(r_0)$ correction; it seems that it does not have a symmetry origin, but is related to the integer value of the number of spatial dimensions. 

The structure of this paper is as follows.  We start with Sec.~\ref{sec:prelim} where we review the conventional effective field theory description of fermions near unitarity.  We show that this theory can be thought of as a nonrelativistic conformal field theory deformed by one relevant and one irrelevant operator, with the coefficients proportional to $1/a$ and $r_0$.  We then proceed with a systematic analysis of local operators in the theory of unitarity fermions. In Sec.\ \ref{lotcf}, we provide a nonperturbative definition of the local operators, and illustrate it on the examples of charge-two and charge-three operators, for which the scaling dimensions can be determined exactly.  We then develop the technique of conformal perturbation theory in Sec.\ \ref{stpfin} and compute the corrections to the two-point functions from the scale-invariance-breaking terms in the Lagrangian in Sec.\ \ref{cptfsar}.  We conclude with Sec.\ \ref{sec:conclusion} and relegate technical details to the Appendices.

\section{Preliminaries: The NRCFT of fermions near unitarity}
\label{sec:prelim}

In this Section, we consider the conventional EFT description of fermions near unitarity.  We then show that this theory can be understood as a NRCFT perturbed by additional operators.

We start with the effective Lagrangian of spin-$1/2$ fermions with point-like interaction\footnote{We set the mass of the single particle state $m=1$ for the rest of the paper.},
\begin{equation}
  \mathcal L = \sum_{\sigma=\uparrow\!\downarrow} \psi^\+_\sigma 
   \left( i \d_t + \frac{\nabla^2}2 \right) \psi_\sigma
  +c \psi^\+_\downarrow \psi^\+_\uparrow 
  \psi_\uparrow \psi_\downarrow .
\end{equation}
Using a Hubbard-Stratonovich transformation, the Lagrangian can be rewritten as 
\begin{equation}\label{L-HS}
  \mathcal L = \sum_\sigma \psi^\+_\sigma 
   \left( i \d_t + \frac{\nabla^2}2 \right) \psi_\sigma
  - \frac1{c} d^\+ d + \psi^\+_\downarrow \psi^\+_\uparrow d
  + d^\+ \psi_\uparrow \psi_\downarrow .
\end{equation}
where $d$ is an auxiliary field. The full inverse propagator of $d$ can be computed by evaluating a fermion one-loop diagram. One finds
\begin{equation}
  D^{-1}(p_0, \vec p) = - \frac1c + \int\! \frac{d^3\vec q}
  {(2\pi)^3}\, \frac{1}{ q^2 + \frac{ p^2}4-p_0-i\epsilon} \,.
\end{equation}
The loop integral diverges linearly in the ultraviolet (UV).  Near unitarity, the inverse four-fermion coupling $1/c$ is fine-tuned to almost cancel the diverging integral, leaving out a finite part:
\begin{equation}\label{dimerD}
  D^{-1}(p_0, \vec p) = - \frac1 {4\pi} \sqrt{\frac{ p^2}4 - p_0 - i\epsilon}
    +\frac1{4\pi a}
\,,
\end{equation}
where $a$ is defined through
\begin{equation}
  \frac1{4\pi a} = - \frac1{c} + \int\!\frac{d^3\q}{(2\pi)^3}\,
  \frac1{q^2} \,,
\end{equation}
and is physically the $s$-wave scattering length.  Once this fine tuning is done, all physical quantities are no longer dependent on the UV cutoff.

We define the critical four-fermion coupling at which the scattering length is infinite as $c_*$,
\begin{equation}
  \frac1{c_*} = \int\!\frac{d^3\q}{(2\pi)^3}\,
  \frac1{q^2} \,.
\end{equation}
The theory at this value of coupling,
\begin{equation}\label{deflag}
  \mathcal L_{\text{CFT}} = \sum_\sigma \psi^\+_\sigma 
   \left( i \d_t + \frac{\nabla^2}2 \right) \psi_\sigma
  - \frac1{c_*} d^\+ d + \psi^\+_\downarrow \psi^\+_\uparrow d
  + d^\+ \psi_\uparrow \psi_\downarrow,
\end{equation}
is a nonrelativistic conformal field theory.  
In this theory, $d$ is a local operator. The two-point Green's function of $d$, in coordinate space, is
\begin{equation}
  \< T d(t, \x) d^\+(0, \vec 0) \> =-\frac{4 }{ \pi  t^2} \exp\left(i\frac{ x^2}{t}\right).
\end{equation}
This has the form of the two-point function of a primary operator with scaling dimension $\Delta_d=2$.

Now when $c$ is away from the finely tuned value, the Lagrangian can be written as
\begin{equation}
  \mathcal L = \mathcal L_\text{CFT} 
  + \frac1{4\pi a}d^\+ d .
\end{equation}
This means that the theory describing fermions with finite scattering length, zero-range interaction is the conformal theory, deformed by the operator $d^\+d$.  This operator has dimension $2\Delta_d=4$ and is hence relevant (recall that in a nonrelativistic theory in 3 spatial dimensions, the dimension of a marginal operator is 5).

To describe an interaction with a finite range, one needs to include into the theory irrelevant interactions.  Including the irrelevant term with lowest dimension (6) leads to the Lagrangian
\begin{equation}\label{d-pert}
  \mathcal L = \mathcal L_\text{CFT} 
  + \frac1{4\pi a}d^\+ d
  -\frac{r_0}{8\pi} d^\+ \left( i \overset{\leftrightarrow}{\d}_t + \frac{\overrightarrow{\nabla}^2+\overleftarrow{\nabla}^2}4\right) d.
\end{equation}
The last term is the only dimension-6 operator which is Galilean invariant. By computing the amplitude of the $s$-wave scattering of two $\psi$-particles, one can identify $r_0$ with the effective range.

\section{Local operators and their correlation functions}\label{lotcf}
 In this section we introduce how local operators are defined in the NRCFT and illustrate the formalism on the example of the charge-two operator.

\subsection{Hilbert space and Hamiltonian of NRCFT}

Here we provide an alternative nonperturbative definition of the NRCFT which will turn out to be convenient for our future calculations.

The Hilbert space of a non-relativistic system is decomposed into sectors with different number of particles
\begin{equation}
  \mathcal H = \mathcal H_0 \otimes \mathcal H_{1,0} \otimes \mathcal H_{0,1} \otimes
  \mathcal H_{1,1} \otimes \mathcal H_{2,1} \otimes \cdots
\end{equation}
where $\mathcal H_{m,n}$ is a sub-Hilbert space of states with $m$ spin-up and $n$ spin-down fermions.

The states in the sector $\mathcal H_{m,n}$ are parameterized by the wave functions
\begin{equation}
  \Psi(\{\x_i; \y_j\}) \equiv \Psi(\x_1, \x_2, \ldots, \x_m; \y_1, \y_2, \ldots, \y_n) .
\end{equation}

Fermi statistics requires that $\Psi$ is antisymmetric under the exchange of any pair of $x$'s or $y$'s. The unitarity $s$-wave interaction implies a boundary condition (the Bethe-Peierls boundary condition) on the behavior of the wave function when a spin-up and a spin-down particle approach each other.  One requires that the $s$-wave part of the wave function is singular, and expandable in a series over odd powers of $|\x_i-\y_j|$ starting from power $-1$:
\begin{equation}\label{BP}
   \Psi(\x_i, \y_j) \to \frac{C_0\left(\frac{\x_i+\y_j}2\right)}{|\x_i-\y_j|}  + |\x_i-\y_j|  C_1\left(\frac{\x_i+\y_j}2\right) + |\x_i-\y_j|^3  C_2\left(\frac{\x_i+\y_j}2\right) + \cdots 
   + \text{non-$s$-wave parts,}
\end{equation}
where we have suppressed the dependence on the coordinates other than $\x_i$ and $\y_j$. 

The Hamiltonian is the sum of the kinetic energies of all particles,
\begin{equation}
  H = -\frac12 \sum_{i=1}^m \frac{\d^2}{\d\x_i^2}
  -\frac12 \sum_{j=1}^n \frac{\d^2}{\d\y_j^2}\,.
\end{equation}
The Hamitonian does not take a state out of the space of wave functions~(\ref{BP}) and is a Hermitian operator.

\subsection{Charge-two operator}
\label{sec:charge-two}

We will now define, within the Hilbert-space approach, the charge-two operator $d$ in the field theory of Sec.~\ref{sec:prelim}.  Let us first limit ourselves to the charge-two sector of the Hilbert space, which contains the two particle states which can be expressed as,  
\begin{equation}
    |\Psi \rangle =     \int_{\vec x, \vec y}
    \Psi (\Vec{x},\Vec{y})  \psi^\dagger_\downarrow(\Vec{y})\psi^\dagger_\uparrow(\Vec{x})|0\rangle .
\end{equation}
The wave function $\Psi(\Vec{x},\Vec{y})$ satisfies the Bethe-Pierls boundary condition~(\ref{BP}) with infinite scattering length to account for the zero-range interaction.

We now define a charge-two operator $O_2(\vec X)$ by giving its matrix elements:
\begin{equation}\label{O2matrixel}
\langle 0| O_2(\vec X) |\Psi \rangle = \lim_{\substack{ \Vec{x} \to \Vec{X} \\ \vec y \to \vec X}} |\Vec{x}-\Vec{y}| \Psi(\Vec{x},\Vec{y}) ,
\end{equation}
where the factor $|\Vec{x}-\Vec{y}|$ ensures that the matrix elements are finite despite the singular boundary condition. Local operators in NRCFTs have well defined quantum numbers---mass (or particle number, if all particle types carry the same mass) and scaling dimension \cite{Nishida_2007}. The particle number of our charge-two operator is fixed by virtue of its definition: $N_{O_2}=-2$, while the scaling dimension can be worked out by the following alternative definition of the charge-two operator in terms of an equal-time operator product expansion
\begin{equation}\label{O2OPEdef}
    \psi_\uparrow(\vec x)\psi_\downarrow(\vec y)\sim \frac{1}{|\vec x-\vec y|} O_2\left(\frac{\vec x + \vec y}{2}\right) + \cdots .
\end{equation}
Considering the fact that the scaling dimension of the fermion field is $\Delta_{\psi}=\frac{3}{2}$ (in $d$ dimensions $\Delta_{\psi}=\frac{d}{2}$) , we get, $\Delta_{O_2}=2 \Delta_{\psi}-1=2$. We can also check that the operator defined in Eq.~\eqref{O2OPEdef} is a primary operator, i.e., satisfies   
\begin{equation}
  [K_i, \, O_2(\vec 0)] = [C, \, O_2(\vec 0)] = 0.
\end{equation}
Where $K_i$ and $C$ are the operators of Galilean boost and proper conformal transformation, respectively.

We will now evaluate the two-point correlation functions of $O_2$ and connect this operator with the operator $d$ in Sec.~\ref{sec:prelim}.

\subsubsection{Two-point Green's function of charge-two operator in NRCFT}
We will be interested in the time-ordered Green's function of $O_2$, 
\begin{equation}
    G_{O_2}(t,\vec x) = -i \Theta(t)\langle 0| O_2(t,\vec x) O^\dagger_2(0,\vec 0) |0 \rangle .
\end{equation}
The time-ordered Green's function coincides with the retarded Green's function thanks to $O_2(t,\vec x)|0\rangle=0$. 

In order to compute the CFT two point function, we insert complete set of states between the operators $O_2$ and $O_2^\dagger$,  \begin{equation}\label{2ptO2cft}
    \langle 0 |O_2 (t,\vec x) O_2^\dagger(0, \vec 0) |0\rangle 
    = \sum_{n}\langle 0| O_2 (t,\vec x) | n\rangle \langle n |O_2^\dagger(0, \vec 0) |0 \rangle .
\end{equation}
Because of number conservation, only two-particle states contributes to the sum.\footnote{Recall that in a relativistic CFT, such a simlification does not occur and we have to take into account multiparticle states as well: relativistic theories contain both particles and holes.} 
Thus, in order to evaluate this sum, we need to count all the two-particle intermediate states. This is possible in our theory because all solutions to the Schr\"odinger equation with Bethe-Peierls boundary condition are known.

In addition, the operator $O_2$ has nonvanishing matrix elements only for states with zero angular momentum between the two particles and the vacuum.  In three spatial dimensions, the Schr\"odinger equation for the two-particle states with zero relative angular momentum is 
\begin{equation}
    \left(\frac{1}{4} \vec\nabla^2_{R_{\rm cm}} + \frac{\partial^2}{ \partial^2 r} + \frac{2}{r}\frac{\partial}{\partial r}\right) \Psi(\vec R_{\rm cm},\vec r ) = - E \Psi(\vec R_{\rm cm}, \vec r), 
\end{equation}
where 
\begin{equation}
    \vec{R}_{\rm cm}=\frac{\vec{x}+\vec{y}}{2},\qquad\vec{r}=\vec{x}-\vec{y},
\end{equation}    
are the center-of-mass and relative coordinates, respectively.

The energy eigenstates are labeled by the center-of-mass momentum $\vec P_{\rm cm}$ and the relative momentum $k$,
\begin{equation}
  \Psi_{\vec P_{\rm cm}, k}(\vec R_{\rm cm}, \vec r)
  = C_{\vec P_{\rm cm}, k} 
   \frac{\cos kr}{r}
  e^{i \vec P_{\rm cm} \cdot \vec R_{\rm cm}},
  \qquad
  E_{\vec P_{\rm cm}, k}= \frac{P_{\rm cm}^2}{4}+ k^2,
\end{equation}
where $C_{\vec P_{\rm cm}, k}$ is the normalization factor and $r= |\vec r|$.  Note that the Bethe-Peierls boundary condition has already been taken into account; the alternative solution to the radial wave equation $r^{-1}\sin kr$ was rejected.

To find the normalization factor $C_{\vec P_{\rm cm}, k}$ and to count the states, we can assume the center-of-mass coordinate $\vec R_{\rm cm}$ runs in a rectangular box of volume $V$ with a periodic boundary condition, while the relative coordinate $r$ is confined in a finite sphere of radius $R_\text{max}$ with the hard-wall boundary condition $\psi(R_\text{max})=0$, which quantizes $k$ to discrete values
\begin{equation}
  k=\frac{(2n +1)\pi}{2 R_\text{max}}, \quad n \in \mathbb{Z}.
\end{equation}
The normalization coefficient can now be computed, and we have
\begin{equation}\label{charge2wavefn}
    \Psi_{\vec P_{\rm cm}, k} (\vec R_{\rm cm}, \vec r)= \frac{e^{i \vec P_{\rm cm} \cdot \vec R_{\rm cm}}}{\sqrt{2 \pi R_\text{max} V}}\frac{\cos (k r)}r .
\end{equation}

At large $V$ and $R_\text{max}$ the sum over intermediate states can be transformed into integrals using the usual rules,
\begin{equation}\label{2bodycontlimit}
 \sum_{\vec P_{\rm cm}}\rightarrow \frac{V}{(2\pi)^3} \int\! d^3\vec P_{\rm cm},\qquad \sum_{k} \rightarrow \frac{R_\text{max}}{\pi} \int\limits^\infty_0\! dk .
\end{equation}

The two-point function of the charge-two operator  in Eq.~\eqref{2ptO2cft} takes the form
\begin{equation}
\langle 0| O_2 (t, \vec x) O_2^\dagger(0, \vec 0) |0 \rangle = \frac{V R_\text{max}}{(2\pi)^3\pi} \int\! d^3\vec P_{\rm cm}\! \int\limits_0^\infty\! dk\, \langle 0| O_2 (t,\vec x) | \Psi_{\vec P_{\rm cm}, k}\rangle \langle \Psi_{\vec P_{\rm cm}, k} |O_2^\dagger(0,\vec 0)|0 \rangle .
\end{equation}
We use the definition of the matrix elements of the charge-two operator as defined in Eq.~\eqref{O2matrixel}, as well as 
Heisenberg evolution, $O(t,\vec x)=e^{i H t} O(0,\vec x)e^{-i H t}$ to get, 
\begin{equation}\label{undefo2}
\langle 0| O_2 (t,\vec x) O_2^\dagger(0, \vec 0) |0 \rangle = \frac{1}{(2\pi)^4\pi} \int\! d^3 \vec P_{\rm cm}\int\limits_0^\infty\! dk\, \exp\left(i \vec P_{\rm cm} \cdot \vec{x} - \frac i4 P_{\rm cm}^2 - ik^2 t\right) .
\end{equation}
Using the integrals in Appendix~\ref{usefulint}, we find
\begin{equation}\label{2charge2ptfn}
  G_{O_2}(t, \vec{x}) = -i \Theta(t)\langle 0| O_2 (t, \vec x) O_2^\dagger(0,\vec 0) |0 \rangle=  \frac{i\Theta(t)}{4 \pi ^3 t^2 }\exp{\left(i \frac{x^2}{t}\right)}\,.
\end{equation}
To compute this Green's function in momentum space, we can use the general formula,
\begin{equation}\label{ft}
 \int\! dt\, d^3\Vec{x}\, \frac{\Theta(t)}{t^\Delta} \exp \left( i \frac{M x^2}{2 t}+ i \omega t -i \Vec{p}\cdot\Vec{x} \right) = i^{\Delta-1} \left(\frac{2 \pi}{M}\right)^{3/2} \left( \frac{ p^2}{2M}- \omega\right)^{\Delta-5/2} \Gamma\left(\frac52 - \Delta\right),  
\end{equation}
to get 
\begin{equation}\label{gcfto2}
G_{O_2}(\omega, \vec{p}) =  -\frac{1}{4\pi \sqrt{\frac{p^2}{4}- \omega}} \,.
\end{equation}
Comparing with Eq.~\eqref{dimerD} in the large scattering length limit, we identify the dimer field with the $O_2$ operator that we have defined in Sec.~\ref{sec:prelim},
\begin{eqnarray}\label{dimero2norm}
    O_2 = \frac{d}{4\pi}\,.
\end{eqnarray}

\subsubsection{Correlation functions and deformations}

We will be interested in the correlation functions when the theory is deformed from the conformal fixed point.  For the problem of fermions at unitarity, the most important deformations
are given by the the inverse scattering length $a^{-1}$ and the effective range $r_0$, and the Lagrangian of the perturbation is, according to Eqs.~(\ref{d-pert}) and (\ref{dimero2norm})
\begin{equation}\label{Lpert}
  \mathcal L_{\rm pert} = \frac{4\pi}a O_2^\dagger O_2 - \pi r_0 O_2^\dagger \left( i\overset{\leftrightarrow}{\d}_t + \frac{\overrightarrow{\nabla}^2+ \overleftarrow{\nabla}^2}4\right) O_2.
\end{equation}
This perturbation Lagrangian contains two operators of lowest dimension: a relevant operator of dimension 4, multiplied by $a^{-1}$, and the leading irrelevant operator of dimension 6, whose coefficient is $r_0$. 

In relativistic theory, there is a powerful technique of conformal perturbation theory. We will develop this technique for the problem at hand.  The idea is to view the NRCFT as the starting point and develop perturbation theory over $1/a$ and $r_0$. This bypasses the need for ``resummation'' of perturbative series in the conventional effective-field-theory approach.

We are interested in computing time-ordered correlation functions of $k$ local operators $\mathcal O_i$, $i=1,\ldots,k$ in the perturbed CFT.  This correlator can be formally written as a path integral, 
\begin{equation}
G_{\mathcal O_1,\ldots\mathcal O_k}
(t_1,\vec x_1; \ldots; t_k, \vec x_k) = \int \mathfrak{D}\psi \,\mathfrak{D} d~ \prod^{k}_{i=1} \mathcal O_i(t_i,\vec x_i)  e^{i\!\int\! dt\, d\vec x\,\mathcal{L}} .
\end{equation}
Writing $\mathcal L = \mathcal L_{\rm CFT}+\mathcal L_{\rm pert}$ and expanding over $\mathcal L_{\rm pert}$, the correlator can be written as
\begin{equation}
 G_{\mathcal O_1,\ldots\mathcal O_k}
(t_1,\vec x_1; \ldots; t_k, \vec x_k) = \sum_{n=0}^\infty \frac1{n!}\langle 0| T \prod^{k}_{i=1} \mathcal O_i(t_i,\vec x_i)\left( \int\!dt\,d\vec x\, \mathcal{L}_{\rm pert}(t,\vec x)\right)^n |0 \rangle_{\rm CFT} ,
\end{equation}
where $T$ denotes time ordering and $\mathcal{L}_{\rm pert}$ is given in Eq.~\eqref{Lpert}. 

The main focus in this paper is to study the two-point functions of primary operators and its leading corrections in the inverse scattering $1/a$ and effective range $r_0$. We first illustrate our technique on the example of the two-point function of the charge-two operator which was  defined in Eq.~\eqref{O2matrixel}.  The correlator of $O_2$ is known to all orders in $a^{-1}$ and $r_0$,
\begin{equation}\label{GO2exact}
   G_{O_2}(\omega, \vec p) = \frac1{4\pi}\left[ - \sqrt{\frac{p^2}4 -\omega} + \frac1a + \frac{r_0}2\left(\frac{ p^2}4-\omega \right) \right]^{-1} .
\end{equation}
Our aim here is more modest: to compute the corrections linear in $a^{-1}$ and $r_0$ to $G_{O_2}$:
\begin{eqnarray}\label{o2deformation}
    G_{O_2}= G^{\rm CFT}_{O_2} + \frac{4\pi i}a G^\text{rel}_{O_2} - {i\pi r_0} G^{\text{irrel}}_{ O_2} ,
\end{eqnarray}
where $G^{\rm CFT}_{O_2}$ is the two point function in the unperturbed NRCFT and the perturbations to linear orders in $a^{-1}$ and $r_0$ are given by \eqref{Lpert}, 
\begin{subequations}\label{2ptdeformationdef}
\begin{align}
G^{\rm CFT}_{O_2}(t,\x)&= -i\langle 0|  TO_2(t,\x) O^\dagger_2(0,\vec 0)|0 \rangle , \\
  G^\text{rel}_{O_2}(t,\x) &= - i \!\int\! dt'\, d^3\vec y\, \langle 0| T O_2(t,\vec{x}) O^\dagger_2(0,\vec 0) O_2^\dagger(t',\vec{y}) O_2(t',\vec{y})|0\rangle , \\
  G^{\text{irrel}}_{O_2}(t, \x)&= - i\!\int\! dt'\, d^3\vec y \,
  \langle 0|T O_2(t,\vec{x}) O^\dagger_2(0,\vec 0) O_2^\dagger(t',\vec{y})\biggl( i\overset{\leftrightarrow}{\partial}_{\!t'} + \frac{\overrightarrow{\nabla}_{\!y}^2+\overleftarrow{\nabla}_{\!y}^2}{4} \biggr)O_2(t',\vec{y})|0\rangle .
\end{align}
\end{subequations}

\subsubsection{Corrections to two-point function of charge-two operator}
We now proceed to evaluate the deformed two-point functions. Taking into account $O_2|0\rangle =0$ and $\langle 0|O_2^\dagger=0$, the two deformations can be rewitten as
\begin{subequations}
\begin{align}
   G^\text{rel}_{O_2}(t,\x) &= - i \Theta(t)\!\int\limits_0^t\! dt'\! \int\! d^3\y\, \langle 0| O_2(t,\vec{x}) O^\dagger_2(t',\vec y) O_2(t',\y)O^\dagger_2(0, \vec 0)|0\rangle ,\\ 
    G^{\text{irrel}}_{O_2}(t,\x)&= - i\Theta(t)\!\int\limits^t_0\! dt'\! \int\! d^3\y\, \langle 0| O_2(t,\vec{x}) O^\dagger_2(t',\vec{y})\left( i\overset{
    \leftrightarrow}{\partial}_{\!t'} + \frac{\overrightarrow{\nabla}_{\!y}^2+\overleftarrow{\nabla}_{\!y}^2}{4} \right)O_2(t',\vec{y})O_2(0,\vec 0)|0 \rangle. \label{Girrel}
\end{align}
\end{subequations}
We apply the method previously used to compute the two-point function to evaluate these deformations. We insert a complete set of eigenstates, and note that number conservation constrains that only the vacuum state needs to be inserted between the operator $O_2^\+(t',\y)$ and $O_2(t',\y)$.  For the relevant ($\sim a^{-1}$) deformation we find
\begin{equation}
   G^{\text{rel}}_{O_2}(t,\vec x) = - i \Theta(t)\!\int\limits^t_0\! dt'\! \int\! d^3\y\, \langle 0| O_2(t,\vec{x}) O^\dagger_2(t',\vec{y})|0\rangle \langle 0| O_2(t',\vec{y})O^\dagger_2(0,\vec 0)|0\rangle .
\end{equation}
This can be written as
\begin{equation}
  G^\text{rel}_{O_2}(t, \x) = i\! \int\limits_{-\infty}^\infty\!dt'\!\int\!d^3\y\, G_{O_2}(t-t',\x-\y)G_{O_2}(t',\y),
\end{equation}
which, in momentum space, is simply
\begin{equation}
  G^\text{rel}_{O_2}(\omega,\p) = i[G_{O_2}(\omega,\p)]^2 = \frac i{16\pi^2}\frac{1}{\frac{p^2}4-\omega} \,.
\end{equation}

The irrelevant deformation~(\ref{Girrel}) can be dealt with similarly.  Inserting the only possible intermediate state---the vacuum state---between $O_2^\+(t',\y)$ and $O_2(t',\y)$, we can write
\begin{equation}
  G^\text{irrel}_{O_2}(t, \x) = i\! \int\!dt'\, d^3\y\, G_{O_2}(t-t',\x-\y) \left( i\overset{\leftrightarrow}{\d}_{\!t'}+ \frac{\overrightarrow{\nabla}_{\!y}^2+\overleftarrow{\nabla}_{\!y}^2}{4} \right) G_{O_2}(t', \y) .
\end{equation}
In momentum space, this reads
\begin{equation}
  G^\text{irrel}_{O_2}(\omega, \p)= -2i\left(\frac{p^2}4-\omega \right) G^2_{O_2}(\omega,\p) = -\frac i{8\pi^2}\,.
\end{equation}

To summarize, the two-point function of $O_2$, with the first-order corrections in $a^{-1}$ and $r_0$, is, according to Eq.~(\ref{o2deformation})
\begin{equation}\label{o2finalans}
  G_{O_2}(\omega,\p) = - \frac1{4\pi}\frac1{\sqrt{\frac{p^2}4-\omega}} - \frac1{4\pi a}\frac1{\frac{p^2}4-\omega} - \frac{r_0}{8\pi} \,.
\end{equation}
which can be checked to coincide with Eq.~(\ref{GO2exact}) when the latter is expanded to linear order in $a^{-1}$ and $r_0$.

\subsection{Charge-three operators}

We now turn our attention to the charge-three local operators. In total analogy with the charge-two operator considered in the previous subsection, the most straightforward way to construct a local charge-three operator is by defining its matrix elements between an arbitrary three-particle state of the interacting theory and the vacuum. The central object of our study is therefore the three-body wave functions which solve the free Hamiltonian, with Bethe-Peierls boundary condition to account for the short-range interaction.

The spectrum of dimensions of the charge-three operators crucially depend on the allowed short-distance behaviors of the solutions of the three-particle wave functions.  When the total spin of the three particles is $3/2$, the wave function is totally antisymmetric in coordinate space and the Bethe-Peierls boundary condition plays no role.  As a result, all spin-$3/2$ charge-three operators behave exactly like in the free theory.  In the channel with total spin $1/2$, we have, e.g., two spin-up and one spin-down particles, and the Bethe-Peierls boundary condition can no longer be neglected.

Let us denote the coordinates of the particles as $\x_i$, $i=1,2,3$, and let particles 1 and 3 carry spin pointing upward and particle 2 carry spin pointing downward.  The three-body wave function $\Psi(\x_1, \x_2, \x_3)$ is antisymmetric under the exchange of two spin-up particles: $\Psi(\x_1, \x_2, \x_3)=-\Psi(\x_3,\x_2,\x_1)$.  The wave function blows up as inverse power of the distance when $\x_2$ approaches either $\x_1$ or $\x_3$.  The Bethe-Peierls boundary condition does not, however, directly prescribe the behavior of the wave function when all the coordinates of all three particles shrink to a point.  To determine that, we need to carefully look at the solutions to the three-body Schr\"odinger equation.

In our case, the three-body problem can be completely solved analytically.  For completeness, we present here Efimov's solution, following closely the presentation of Ref.~\cite{werner:tel-00285587}. The Schr\"odinger equation is given by  
\begin{equation}\label{3body-free-H1}
    -\frac{1}{2} \biggl(\sum^3_{i=1}\nabla^2_{\vec{x}_i}\biggr) \Psi(\vec{x}_1, \vec{x}_2, \vec{x}_3) = E \Psi(\vec{x}_1, \vec{x}_2, \vec{x}_3) ,
\end{equation}
and is supposed to be valid when the none of the three coordinates coincide.  We now introduce the Jacobi coordinates,
\begin{subequations}\label{Jacobi}
\begin{align}
  \vec R_\text{cm} &= \frac{\vec{x}_1+\vec{x}_2+\vec{x}_3}{3},\\
  \vec{r} &= \vec{x}_2-\vec{x}_1,\\ \vec{\rho} &= \frac{2}{\sqrt{3}}\left( \vec{x}_3- \frac{\vec{x_1}+\vec{x_2}}{2} \right).
\end{align}
\end{subequations}
For future reference, the Jacobian of the change of coordinates is
\begin{equation}\label{JacobiJacobian}
  \frac{\d(\x_1,\x_2,\x_3)}{\d(\vec R_\text{rm},\vec r,\vec \rho)} = \biggl( \frac{\sqrt3}2\biggr)^3  .
\end{equation}
Assuming the wave function is a plane wave in the center-of-mass coordinate, $\Psi(\vec R_\text{cm}, \vec{r}, \vec{\rho})= \exp(i\vec P_\text{cm}\cdot\vec R_\text{cm}) \psi^{(3)}(\vec{r}, \vec{\rho})$, in the relative coordinates $\vec r$ and $\vec\rho$ the Schr\"odinger equation is
\begin{equation}\label{Schr-rrho}
  -(\nabla^2_{\!r} + \nabla^2_{\!\rho})) \psi^{(3)}(\vec{r}, \vec{\rho}) = k^2 \psi^{(3)}(\vec{r}, \vec{\rho}).
\end{equation}
where the energy of the relative motion is denoted as $k^2$. The total energy is the sum of the energies of the center-of-mass motion and relative motion: $E=\frac16 P_\text{cm}^2+k^2$. 

Equation~(\ref{Schr-rrho}) should be supplemented by the Bethe-Peierls boundary condition at $\x_2\to \x_1$ and $\x_2\to\x_3$.  We defer the detail discussion of this boundary condition to later.

To solve the Schr\"odinger equation, we go from $\vec r$ and $\vec\rho$ (total of six coordinates) to the hyperspherical coordinates, consisting of the hyperradius, 
\begin{equation}\label{hyperradius}
  R= \sqrt{\frac{{r}^2+{\rho}^2}{2}}
  =\sqrt{\frac{(\x_1-\x_2)^2+(\x_2-\x_3)^2+(\x_3-\x_1)^2}3} \,,
\end{equation}
and five hyperangles $\Omega=(\alpha,\hat r,\hat \rho)$, where
\begin{equation}\label{hyperangles}
\alpha=\arctan \frac{r}{\rho}\, ,
\quad \hat{r} = \frac{\vec r}{r}\, ,\quad \hat{\rho} = \frac{\vec\rho}\rho \,.
\end{equation}
Again for future reference, we note the Jacobian of the coordinate change,
\begin{equation}\label{EfimovJacobian}
  d^3\vec r\, d^3\vec\rho = 2R^5 \sin^2 2\alpha\, dR\, d\alpha\, d^2\hat r\, d^2\hat\rho .  
\end{equation}
In the hyperspherical coordinates, the Schr\"odinger equation has the form
\begin{equation}
  -\frac12 \left( \frac{\d^2}{\d R^2} + \frac5R \frac\d{\d R} + \frac{\nabla^2_{\!\Omega}}{R^2}\right)\psi^{(3)}(R, \Omega) = k^2 \psi^{(3)}({R,\Omega}),
\end{equation}
where the hyperangular Laplace operator is
\begin{equation}\label{hyperangLaplace}
  \nabla^2_{\!\Omega} = \frac{\d^2}{\d\alpha^2} + \frac4{\tan 2\alpha}\frac\d{\d\alpha} + \frac{\nabla^2_{\!\hat r}}{\sin^2\alpha} + \frac{\nabla^2_{\!\hat \rho}}{\cos^2\alpha}\,.   
\end{equation}
The task of finding the solution to the Schr\"odinger equation boils down to finding the spectrum of the hyperangular Laplace operator $\nabla_{\!\Omega}^2$.  Suppose we know the complete spectrum of this operator. We denote the eigenvalues of this operator by $-s^2+4$:
\begin{equation}\label{hyperangLaplace-eq}
  \nabla^2_{\!\Omega}\Phi_s(\Omega) = (-s^2+4)\Phi_s(\Omega).
\end{equation}
Then the solution to the Schr\"odinger equation can be searched in the form
\begin{equation}
  \psi^{(3)}(R,\Omega) = \frac{F_s(R)}{R^2} \Phi_s(\Omega) ,
\end{equation}
where $F_s(R)$ satisfies the hyperradial wave equation
\begin{equation}\label{hyperrad-eq}
  -F''(R)-\frac{1}{R}F'(R)+\frac{s^2}{R^2} =2k^2F(R).
\end{equation}

\subsubsection{Solving the hyperradial wave equation}

We will see that for the problem of spin-$1/2$ fermions, $s^2>0$, so we will consider $s$ real and positive.  There are two independent solutions to the the hyperradial wave equation~(\ref{hyperrad-eq}), $F(R)\sim J_s(\sqrt{2}kR)$ and  $F(R)\sim Y_s(\sqrt{2}kR)$, where $J_s(x)$ and $Y_s(x)$ are the Bessel functions of first and second kind, respectively.  The asymptotics of these solutions at small $R$ are $J_s(\sqrt2kR)\sim R^s$ and $Y_s(\sqrt2kR)\sim R^{-s}$. The normalization integral for the wave function contains a factor $\int\! R^5 dR |F(R)/R^2|^2= \int\! RdR\, |F(R)|^2$. This factor never diverges for the $J_s$ solution, but diverges for the $Y_s$ solution when $s\ge1$.  As we will see, in our case all eigenvalues of the hyperangular Laplacian has $s>1$, so we set

\begin{equation}\label{hyperrad3}
    F(R)=J_s(\sqrt{2}kR).
\end{equation}
In analogy with the charge-two case, we impose the vanishing boundary condition for the wave function at a maximal hyperradius $R_\text{max}$:
\begin{equation}\label{quant3bdyk}
  F(R_\text{max})=0
 \implies k=
 \frac\pi{\sqrt2 R}\left(n+\frac s2+\frac34 \right)
 , \quad n\in \mathbb{Z}.
\end{equation}

We now proceed to define charge-three operators.  First we note that a general three-body state will have the following behavior at short distances,
\begin{equation}\label{expansion-s}
 \Psi(\x_1, \x_2, \x_3) = \sum_s R^{s}\Phi_s(\Omega) \Psi_s(\vec R_\text{cm}),
\end{equation}
when the three coordinates $\x_1$, $\x_2$, $\x_3$ shrink to the same point $\vec R_\text{cm}$ in an uniform way (i.e., with $R\to0$ and all hyperangles fixed).  One can then define the charge-three operators $O_3^s$ so that 
\begin{equation}
  \<0| O_3^s(\vec R_\text{cm}) |\Psi\>
  = c_s \Psi_s(\vec R_\text{cm}) ,
\end{equation}
where $c_s$ is some constant chosen for convenience.
If there are several eigenstates of the hyperangular Laplacian with the same eigenvalues (for example in the cases of states with nonzero angular momentum) then in Eq.~(\ref{expansion-s}) one needs to include in the expansion~(\ref{expansion-s}) the complete orthonormal set of basis eigenvectors $\Phi_s(\Omega)$ with that value of $s$ and define one charge-three operator for each eigenmode (which would form a $SO(3)$ rotational multiplet if the degeneracy is due to rotational symmetry).

For completeness, here we note two other possible behaviors of the three-body sector.
\begin{itemize}
\item When $s^2<0$, the hyperradial wave equation~(\ref{hyperrad-eq}) corresponds to the situation of ``falling to the center.'' The three-body problem is not well-defined without a UV cutoff.  With an UV cutoff, solving Eq.~(\ref{hyperrad-eq}) one finds an infinite number of bound states with binding energies $B_n$ tending to 0 exponentially with the radial quantum number $n$: $B_n\sim e^{-2\pi ns}$.  This is the Efimov effect and is the situation with three identical bosons with unitarity interaction.
\item When $0<s^2<1$, both $J_s$ and $Y_s$ are normalizable at small $R$.  In this case, there are two CFTs, one with non-fine-tuned three-body interaction and the other with a fine-tuned three-body resonance.  The radial wave function $F(R)$ behaves as $R^s$ and $R^{-s}$ at small $R$ in these two CFTs. This is the situation with fermions when the mass difference of the spin-up and spin-down species is within a certain range.
\end{itemize}

\subsubsection{Solving the hyperangular wave equation}\label{sec:solving_hyperangular}

Now we find the spectrum of the Laplace operator in the hyperangular coordinates.  The hyperangular Laplace equation, Eqs.~(\ref{hyperangLaplace}) and (\ref{hyperangLaplace-eq}) look simple, but we need to remember that the wave function has to be antisymmetric under the exchange $\x_1\to\x_3$ and satisfy the Bethe-Peierls boundary condition when $\x_2\to\x_1$ and $x_2\to x_3$.  Since the hyperradius~(\ref{hyperradius}) is invariant under the exchange of any pair of coordinates, Fermi statistics imposes an antisymmetry condition on the hyperangular wave function $\Phi_s(\Omega)$.  The Bethe-Peierls boundary condition when $\x_2\to\x_1$ is, 
\begin{equation}
 \lim_{\vec{r}\rightarrow 0}\psi^{(3)}(\vec{r}, \vec{\rho}) = \frac{1}r A(\vec \rho) + O(r), \qquad r\equiv |\vec r|,
\end{equation}
which can be formulated as a differential constraint,
\begin{equation}\label{BP3bodyrrho}
    \frac{\partial (r \psi^{(3)})}{\partial r}\biggl|_{r=0} =0.
\end{equation}
and has a simple form in the hyperangular coordinates,
\begin{equation}
  \frac{\d (\alpha\Phi(\alpha,\hat r,\hat\rho))}{\d\alpha}\biggl|_{\alpha=0} =0 ,
\end{equation}
but the boundary condition when $\x_2\to\x_3$ is quite complicated when written in the language of the hyperangular coordinates~(\ref{hyperangles}).  The problem is that the Jacobi coordinates~(\ref{Jacobi}) treat the two spin-up particles (particles 1 and 3) unequally.

Efimov's solution makes use of a decomposition of the wave function into the Faddeev components.  We note that Eq.~(\ref{3body-free-H}) is valid only when $\x_2\neq\x_1$ and $\x_2\neq\x_3$; if one wants to write an equation that is valid for all values of the coordinates, one would have delta-functional sources,
\begin{equation}\label{3body-free-H}
    -\frac{1}{2} \sum^3_{i=1}\nabla^2_{\vec{x}_i} \Psi(\vec{x}_1, \vec{x}_2, \vec{x}_3) = E \Psi(\vec{x}_1, \vec{x}_2, \vec{x}_3) + \delta(\x_2-\x_1) F\left(\x_3, \frac{\x_1+\x_2}2\right) -  \delta(\x_2-\x_3) F\left(\x_1, \frac{\x_3+\x_2}2\right),  
\end{equation}
where the function $F$ can be in principle related to $\Psi$ so that Eq.~(\ref{3body-free-H}) becomes a closed equation for $\Psi$, but we will not need this relation.  We can formally write $\Psi$ into the sum of two parts,
\begin{equation}\label{Psi-linear-comb}
   \Psi = \Psi_{12} - \Psi_{32},
\end{equation}
where
\begin{align}
  \Psi_{12} &= \biggl( -\frac{1}{2} \sum^3_{i=1}\nabla^2_{x_i} -E \biggr)^{-1} \delta(\x_2-\x_1) F\biggl(\x_3,\frac{\x_1+\x_2}2\biggr),\label{Psi12}\\
  \Psi_{32} &= \biggl( -\frac{1}{2} \sum^3_{i=1}\nabla^2_{x_i} -E \biggr)^{-1} \delta(\x_2-\x_3) F\biggl(\x_1, \frac{\x_3+\x_2}2\biggr) = P_{13}\Psi_{12},
\end{align}
where the operator $P_{13}$ is the operator that exchanges $x_1$ and $x_3$: $\Psi_{32}(\x_1, \x_2, \x_3)=\Psi_{12}(\x_3, \x_2, \x_1)$. 
$\Psi_{12}$ and $\Psi_{32}$ are called Faddeev components. By construction, they satisfy the free Schr\"odinger equation at non-coinciding coordinates, but they satisfy neither Fermi statistics nor the Bethe-Peierls boundary condition.  Efimov's method is to search for the general expression for the Faddeev components and then impose the Bethe-Peierls constraint by hand [Fermi statistics is built in in the linear combination~(\ref{Psi-linear-comb})].

The Faddeev component $\Psi_{12}$ can also be searched in the form
\begin{equation}
   \Psi_{12} = \frac{F(R)}{R^2} \Phi_{s,12}(\Omega),
\end{equation}
where $\Phi_{s,12}(\Omega)$ satisfies the equation
\begin{equation}
  \nabla^2_{\!\Omega} \Phi_{s,12}(\Omega) = (-s^2+4) \Phi_{s,12}(\Omega),
\end{equation}
and $F(R)$ satisfies Eq.~(\ref{hyperrad-eq}).  As the hyperradius $R$ is symmetric under permutation [see Eq.(\ref{hyperradius})], $\Phi_{s,32}=P_{13}\Phi_{s,12}$.
From Eq.~(\ref{Psi12}) one sees that $\Phi_{12}$ contains only the $s$-wave component in $\hat r$, so one can set in the hyperangular Laplacian $\nabla^2_{\!\hat r}=0$~\footnote{The authors thank F\'elix Werner for explaining this argument, due to Yvan Castin, to them.}.  Searching for the solution of the form
\begin{equation}
  \Phi_{s,12}(\Omega) = \frac{\phi^l_s(\alpha)}{\sin (2\alpha)}Y_l^m(\hat{\rho}),
\end{equation}
where the spherical function is normalized as
\begin{equation}\label{sphharmnorm}
 Y_l^m(\theta, \varphi) =  e^{i m \varphi } \sqrt{\frac{(l-m)!}{(l+m)!}} P_l^m(\cos (\vartheta )), \qquad
 \int\! d\Omega\, |Y_l^m(\theta, \varphi)|^2 = \frac{4\pi}{2 l +1} \,,
\end{equation}
we find the equation for $\phi_s^l(\alpha)$:
\begin{equation}
{\phi^l_s}''(\alpha) - \frac{l(l+1)}{\cos^2 \alpha}\phi^l_s(\alpha)= -s^2 {\phi^l_s} (\alpha).
\end{equation}
The Bethe-Pierls boundary condition \eqref{BP3bodyrrho}, which should be imposed on $\Phi_{s,12}(\Omega)-P_{13}\Phi_s(\Omega)$, and the regularity in $\rho \rightarrow 0$ limit, in these coordinates, translate to the following condition on $\phi(\alpha)$:
\begin{align}
   {\phi'}^l_s(0)-(-1)^l\frac{4}{\sqrt{3}}\phi^l_s(\pi/3)&=0, \label{3bodyhypangbc-1}\\
    \phi^l_s(\pi/2)&=0. \label{3bodyhypangbc-2}
\end{align}
In deriving Eq.~(\ref{3bodyhypangbc-1}) we have made use of the fact that that $r\rightarrow 0$ implies $\alpha \rightarrow 0$, $P_{13}\alpha \rightarrow 
\frac{\pi}{3}$, and $P_{13}\hat\rho \to -\hat\rho$. We will be interested in the cases of $l=0$ and $l=1$. The hyperangular wave functions satisfying the second boundary condition~\eqref{3bodyhypangbc-2} are given by  \begin{equation}\label{hypangw3bdy}
\phi^{l=0}_s(\alpha) =\sin\! \left[s\left(\frac{\pi}{2}-\alpha\right) \right],\qquad
\phi^{l=1}_s(\alpha) = -s \cos\! \left[s \left(\frac{\pi}{2}-\alpha\right)\right] + \tan\alpha\, \sin\! \left[s \left(\frac{\pi}{2}-\alpha\right)\right] ,
\end{equation}
while the first boundary condition becomes an equation determining the spectrum of $s$,
\begin{align}\label{recrelns3d}
  l=0: &\qquad s\cos\left(\frac\pi2 s \right) + \frac4{\sqrt3}\sin\left(\frac\pi6s\right) = 0,\\
  l=1: &\qquad (s^2-1) \sin\left(\frac\pi2 s\right)
  + \frac4{\sqrt3} s\cos\left(\frac\pi6s \right) - 4\sin\left(\frac\pi6s\right)=0   .
\end{align}
(The equation for a general value of $l$ can be found in Ref.~\cite{werner:tel-00285587}.) There is an infinite number of solutions for each of the equations; the numerical values for the lowest four solutions for each $l$ are given in Table~\ref{table1}.  (The two spurious solutions $s=2$ for $l=0$ and $s=1$ for $l=1$ have been ignored, since the both lead to $\Phi_s(\Omega)=0$.)  We also note here the asymptotic behavior of $n$th solution of each equation at large $n$,
\begin{align}\label{sn-asympt}
  s_n = \begin{cases}
    2n+1 + O \left( n^{-1}\right), & l = 0,\\
    2n + O\left( n^{-1} \right), & l = 1.
  \end{cases}
\end{align}
\begin{table}[ht]
	\begin{center}
		\begin{tabular}{|c|c|}
			\hline
			$l=0$ & $l=1$ \\
			\hline
			$~2.16622~$ & $~1.77272~$ \\
                \hline
                5.12735 &  4.35825 \\
                \hline
                7.11448 & 5.71643 \\
			\hline
                8.83225 & 8.05319 \\
          \hline
		\end{tabular}
  \end{center}
\caption{The first four solutions for $s$ for $l=0, 1$.}
  \label{table1}
 \end{table}

Once the function $s$ has been found, the hyperangular part of the wave function can be computed through 
\begin{equation}
  \Phi_s(\Omega) = (1-P_{13}) \frac{\phi^l_s(\alpha)}{\sin (2\alpha)}Y_l^m(\hat{\rho}).
\end{equation}

\subsubsection{Normalizing energy eigenstates}

We normalize the three-body wave function by requiring 
\begin{equation}
  \int\! d^3\x_1\, d^3\x_2\, d^3\x_3 \, |\Psi(\x_1, \x_2, \x_3)|^2 = 1.
\end{equation}
The three-body state is then related to the the wave function by
\begin{equation}
  |\Psi \> = \frac1{\sqrt2}\!\int\!d^3\x_1\, d^3\x_2\, d^3\x_3\, \Psi(\x_1, \x_2, \x_3) \psi_\uparrow^\+(\x_1) \psi_\downarrow^\+(\x_2) \psi_\uparrow^\+ (\x_3) |0\>. 
\end{equation}
The energy eigenstates are written as a function of the Jacobi coordinates as
\begin{equation}\label{3body-E-eigenstate}
\Psi_{s,k}^{l, m}(\vec{R}_{\rm cm}, \vec{r}, \vec{\rho})
 = \left(\frac 2{\sqrt3}\right)^{3/2} N^{l}_k\, e^{i \vec P_{\rm cm}\cdot \vec R_{\rm cm}}\frac{J_s(\sqrt{2} k R)}{R^2} 
 \Phi^{l,m}_s(\Omega),
\end{equation}
where $N^l_k$ is a normalization constant and the factor $(2/\sqrt3)^{3/2}$ has been inserted for convenience.  Using Eqs.~(\ref{JacobiJacobian}) and (\ref{EfimovJacobian}), the normalization integral can be written as
\begin{equation}
  1 = |N^l_k|^2 \int\!d^3\vec R_{\rm cm}\! \int\! RdR \, J_s^2(\sqrt2 kR)\! \int\! d\bar\Omega \, |\Phi^l_s(\Omega)|^2,
\end{equation}
where $d\bar\Omega\equiv 2\sin^2 2\alpha\, d\alpha\, d^2\hat\rho\, d^2\hat r$.  As before, we assume that $\hat R_{\rm cm}$ is confined in a large, but finite box of volume $V$ and the hyperradius has a large, but finite maximal value of $R_{\rm max}$ where the wave function vanishes, the normalization factor is
\begin{equation}
  |N^l_k|^2 = \frac{\sqrt2\pi k}{VR_{\rm max} f_s^l}\,,
\end{equation}
where we have defined
\begin{equation}\label{normfs}
  f^l_s \equiv \int\! d\bar\Omega \, |\Phi^{l, m}_s(\Omega)|^2.
\end{equation}

\subsubsection{Defining charge-three operators}

In analogy with the case of the charge-two operator, we now define charge-three operators by its matrix element between the vacuum and the three particle momentum eigenstate,
\begin{equation}
  \< 0 | O_3^{l, m, s}(\vec R_{\rm cm}) | \Psi^{(3)} \> = \lim_{R\to0} R^{2-s} \int\!d\bar\Omega\, \Psi^{(3)}(\vec R_{\rm cm}, R, \Omega) \Phi_s^{l, m}(\Omega).
\end{equation}
For example, for the normalized energy eigenstate given by the wave function~(\ref{3body-E-eigenstate})
\begin{equation}\label{o3def1}
    \langle 0| O^{l',m',s'}_3(\vec R_{\rm cm})|\Psi^{(3)}_{\vec P_{\rm cm},l,m,s,k}\rangle=\left(\frac 2{\sqrt3}\right)^{3/2} N^l_k f^l_s \frac {k^s}{2^{s/2}\Gamma(s+1)}  e^{i \vec P_{\rm cm}\cdot \vec R_{\rm cm}} \delta_{ss'}\delta_{ll'} \delta_{mm'} \,.
\end{equation}

We see that there is an infinite number of charge-three operators, labeled by the eigenmodes of the three-body hyperangular Laplacian. 

The charge-three operators should appear in the operator product expansion (OPE) of a charge-two and a charge-one operator.  To find the coefficients of this OPE, we compute the matrix element 
\begin{align}\label{O2psi-matrix-el}
    \langle 0| O_2(\y)\psi_{\uparrow}(\x)|\Psi^{(3)}_{\vec P_{\rm cm}, l,m,s,k} \rangle &= \lim_{z \rightarrow 0} |\vec{z}|\langle 0| \psi_{\uparrow}(\y+\frac{\z}{2}) \psi_{\downarrow}(\y-\frac{\z}{2})\psi_{\uparrow}(\x)|\Psi^{(3)}_{\vec P_{\rm cm},l,m,s,k} \rangle \nonumber\\
    &= \sqrt{2}\lim_{|\vec{z}|\rightarrow 0} |\vec{z}|\Psi^{(3)}_{\vec P_{\rm cm}, l, s,k}\left(\frac{2\y+\x}{3}, -\vec{z}, \frac{2}{\sqrt{3}}(\x-\y)\right)\nonumber\\
    &= \sqrt{2}\left(\frac 2{\sqrt3}\right)^{3/2} N^{l}_k \exp\!\left[\frac i3 \vec P_{\rm cm}\cdot (2\y+\x)\right] \frac{J_s(\frac{2}{\sqrt{3}}k|y-x|)}{\frac2{\sqrt3}|y-x|}\phi^{l}_s(0) Y^l_m(\widehat{x-y}).
\end{align}
Comparing with Eq.~(\ref{o3def1}), one finds\footnote{For the interested reader, the the full equal-time OPEs in the charge-three sector (i.e.,  including the descendant contributions) are worked out in Appendix~\ref{opeappendix}.}
\begin{equation}
  O_2(\y) \psi_\uparrow(\x) = \sum_{s,l,m} 
  \left( \frac23 \right)^{\frac{s-1}2} \frac{\phi_s^l(0)}{f^l_s}  Y^l_m(\widehat{x-y}) |\y-\x|^{1-s} O_3^{l,m,s}\left(\frac{2\y+\x}{3}\right)
  + \text{descendants} .
\end{equation}

We now compute the two-point function of the charge-three operator $O^{l,m,s}_3(t,\x)$. We insert complete set of states to get, \footnote{Note that due to orthogonality of the hyperangular wave functions, only the two-point functions of operators with the same $s$, $l$, and $m$ are nonzero.}
\begin{align}
    \langle 0| O^{l,m,s}_3 (t,\x) O^{\dagger, l,m,s}_3 (0,\vec 0)|0 \rangle &= \sum_{\vec P_{\rm cm}, k}\langle 0| O^{l,m,s}_3 (t,\x)|\Psi^{(3)}_{l,m,s,k}\rangle \langle \Psi^{(3)}_{l,m,s,k} | O^{\dagger, l,m,s}_3 (0,\vec 0)|0 \rangle \nonumber\\
    &= \frac{\sqrt{2}V R_\text{max}}{(2\pi)^3 \pi} \int\! d^3 \vec P_{\rm cm}\! \int\limits^\infty_0\! dk\,  \langle 0| O^{l,m,s}_3 (t,\x)|\Psi^{(3)}_{l,m,s,k}\rangle \langle \Psi^{(3)}_{l,m,s,k} | O^{\dagger, l,m,s}_3 (0,\vec 0)|0 \rangle,
\end{align}
where we have used the fact that in the infinite volume limit the sum over discrete eigenvalues go to the continuum obeying the following rules 
\begin{equation}
 \sum_{\vec P_{\rm cm}}\rightarrow \frac{V}{(2\pi)^3}\! \int\! d^3 \vec P_{\rm cm},\qquad \sum_{k} \rightarrow \frac{\sqrt{2}R_\text{max}}{\pi}\! \int\limits^\infty_0\! dk ,
\end{equation}
the latter is due to the spacing between the quantized values of $k$ in Eq.~(\ref{quant3bdyk}).
Using Eq.~\eqref{o3def1}, we find 
\begin{align}
 \langle 0| O^{l,m,s}_3 (t,\x) O^{\dagger, l,m,s}_3 (0,\vec 0)|0 \rangle &= \left(\frac{2}{\sqrt{3}}\right)^3 \frac{2^{1-s} f^l_s}{(2\pi)^3 \Gamma(1+s)^2}\! \int\! d^3\vec P_{\rm cm}\, \exp\left(-\frac{i}6  P_{\rm cm}^2  t+i \vec P_{\rm cm}\cdot \vec x\right)\!\int\limits^\infty_0\! dk\, k^{2s+1} e^{-i k^2 t} \nonumber\\
 &=\frac{2^{-s+\frac{3}{2}}f^l_s }{\Gamma(1+s)\pi^{\frac{3}{2}}} \frac{1}{(i t)^{s+\frac{5}{2}}}
 \exp\left(\frac{3 i x^2}{2t}\right) .
\end{align}
Thus, we get
\begin{equation}\label{o32ptgenl}
 \langle 0| O^{l,m,s}_3 (t,\x) O^{\dagger, l',m',s'}_3 (0,\vec 0)|0 \rangle 
 = \frac{2^{-s+\frac{3}{2}}f^l_s}{\Gamma(1+s)\pi^{\frac{3}{2}}} \frac{1}{(i t)^{s+\frac{5}{2}}}
 \exp\left(\frac{3 i {x}^2}{2t}\right) \delta_{ss'} \delta_{ll'} \delta_{mm'} \,.
\end{equation}
This has the form of a two-point function of a primary operator with mass $M=3$ and dimension $\Delta=s+\frac52$.

\section{Some three-point functions in NRCFT}\label{stpfin}
In this section we evaluate some three point functions that wil be needed later for the conformal perturbation theory. 

\subsection{$\langle O_2\psi_\downarrow^\+\psi_\uparrow^\+\rangle$}
As an warm up exercise, we compute the three point functions of the charge-two operator with two $\psi$ fields.  This calculation will be used in the test case for conformal perturbation theory in Sec.~\ref{testcase}.  We will only need the expression of the three-point function when the two $\psi$ fields are evaluated at the same time.

We evaluate the three point function $\langle 0| O_2\psi_\uparrow^\+\psi_\downarrow^\+ |0\rangle$ using the same quantum mechanical approach that we followed for the two-point function. We insert the complete set of two-particle energy eigenstates to get, 

\begin{equation}
\langle 0| T O_2(0,\vec{0}) \psi_\downarrow^\dagger(t,\vec{y})\psi_\uparrow^\dagger(t,\vec x)|0\rangle =  \Theta(-t)\frac{V R_\text{max}}{(2\pi)^3\pi}\! \int\! d^3\vec P_{\rm cm}\, dk\, \langle 0| O_2(0,\vec{0})| \Psi^{(2)}_{\vec P_{\rm cm}, k}\rangle \langle \Psi^{(2)}_{\vec P_{\rm cm},k}| \psi_\downarrow^\dagger(t,\vec y)\psi_\uparrow^\dagger(t, \vec x)|0\rangle .
\end{equation}
These matrix elements have been already evaluated earlier in this paper. From Eqs.~\eqref{charge2wavefn} and \eqref{O2matrixel}, we obtain, 
\begin{subequations}
\begin{align}
    \langle 0| O_2(0,\vec{0})| \Psi^{(2)}_{\vec P_{\rm cm},k}\rangle &= \frac1{\sqrt{2\pi V R}}\, , \\
    \langle \Psi^{(2)}_{\vec P_{\rm cm},k}| \psi_\downarrow^\dagger(t,\vec{y})\psi_\uparrow^\dagger(t,\vec x)| 0\rangle &= - \frac{\cos(k |\vec x-\vec y|)}{\sqrt{2\pi V R}\, |\vec x-
    \vec y|} \exp\!\left[i \vec P_{\rm cm}\cdot \frac{\vec x+\vec y}{2}+ i\biggl(\frac{ P_{\rm cm}^2}{4}+k^2 \biggr)t\right].
\end{align}
\end{subequations}
The three point function takes the form, 

\begin{equation}\label{gmuo21}
    \langle 0|T O_2(0,\vec{0}) \psi_\downarrow^\dagger(t,\vec{y})\psi_\uparrow^\dagger(t,\vec x)|0\rangle 
    = - \frac{\Theta(-t)}{4\pi^3 t^2|\vec x-\vec y|} \exp \! \left( -\frac i2 \frac{x^2+y^2}t\right),
\end{equation}
where we have used integrals in Appendix \ref{usefulint}. Similarly 

\begin{equation}\label{gmuo22}
\langle 0|T \psi_\uparrow(t,\vec x)   \psi_\downarrow (t,\vec y)O^\+_2(0, \vec 0)|0\rangle = -\frac{\Theta(t)}{4\pi^3 t^{2}|\x-\y|} \exp \! \left( \frac i2 \frac{x^2+y^2}t\right).
\end{equation}

\subsection{$\langle O_3 O_2^\+\psi_\uparrow^\+\rangle$}
In this subsection we evaluate the three point functions that are needed for computing the deformation of the charge-three operator two point function. For brevity we will explicitly derive the result for $l=0$ and will only state the result for general $l$. We insert the complete set of three-particle momentum eigenstates to get 
\begin{multline}
  \langle 0|T O^{l=0,s}_3(0,\vec{0}) O^\dagger_2(t,\vec{y})\psi_\uparrow^\dagger(t,\x)|0\rangle\\
  =\Theta(-t)\frac{\sqrt{2}V R_1}{(2\pi)^3 \pi}\! \int\! d^3 \vec P_{\rm cm}\! \int\limits^\infty_0\! dk\,  \langle 0| O^{l= 0 ,s}_3(0,\vec{0})| \Psi^{(3)}_{\vec P_{\rm cm}, l= 0, s,k}\rangle\langle \Psi^{(3)}_{\vec P_{\rm cm}, l= 0, s,k}| O^\dagger_2(t,\vec{y})\psi_\uparrow^\dagger(t,\vec x)|0\rangle  .
\end{multline}
These matrix elements are evaluated as follows 
\begin{subequations}
\begin{align}
\langle 0| O^{l=0,s}_3(0, \vec{0})|\Psi^{(3)}_{l=0,s,k}\rangle
    &=\left(\frac{2}{\sqrt{3}}\right)^{3/2}N^0_k f^0_s \frac{2^{-s/2} k^s}{\Gamma(s+1)}\,, \\ 
   \langle \Psi^{(3)}_{l= 0, s,k}| O^\dagger_2(t,\vec{y})\psi_\uparrow^\dagger(t,\x)|0\rangle
&= -\frac{2}{3^{1/4}}N^{0}_k \frac{J_s(\frac{2}{\sqrt{3}} k |\vec{y}-\vec{x}|)}{|\vec{y}-\vec{x}|} {\phi^*}_s^{l=0}(0) \exp\left[ i \biggl(\frac{ P_{\text{cm}}^2}{6}+k^2\biggr)t -i \vec P_{\rm cm}\cdot \frac{2\Vec{y}+\Vec{x}}{3}\right],
\end{align}
\end{subequations}
where we have used Eq.~(\ref{O2psi-matrix-el}).
The three point function then simplifies to 

\begin{equation}
    \langle 0|T O^{l=0,s}_3(0,\vec{0}) O^\dagger_2(t,\vec{y})\psi_\uparrow^\dagger(t,\x)|0\rangle =
    N^{l=0}_{3pt}\Theta(-t) \frac{|\y-\x|^{s-1}}{(-t)^{s+\frac52}} \exp\left[-\frac it \biggl( \frac{x^2}2+y^2\biggr)\right],
\end{equation}
where we have used Eq.~\eqref{normfs} and the integrals listed in Appendix \ref{usefulint}. We have also defined 
\begin{equation}\label{eq:Nls_defn}
    N^{l=0}_{3pt} 
\equiv -\frac{i^{-s-\frac{5}{2}} 2^{-\frac{s}{2}+1} 3^{\frac{1-s}{2}} {\phi^*}_s^{l=0}(0)}{\pi ^{3/2} \Gamma (s+1)}\,.
\end{equation}
 
Similar manipulations give us, 

\begin{eqnarray}
     \langle 0|T \psi_\uparrow(t,\x) O_2(t,\vec{y})O^{l=0 ,s\dagger}_3(0,\vec 0)|0\rangle = {N^{l=0}_{3pt}}^* \Theta(t) \frac{|\y-\x|^{s-1}}{(-t)^{s+\frac52}} \exp\left[\frac it \biggl( \frac{x^2}2+y^2\biggr)\right].
\end{eqnarray}

Generalisation to charge-three operators with arbitrary spin $l$ is straightforward and we have, 

\begin{equation}\label{spinlo3o2psi}
    \langle 0|T O^{l,s}_3(0,\vec{0}) O^\dagger_2(t,\vec{y})\psi_\uparrow^\dagger(t,\x)|0\rangle =
    N^{l}_{3pt}\Theta(-t) \frac{|\y-\x|^{s-1}}{(-t)^{s+\frac52}} \exp\left[-\frac it \biggl( \frac{x^2}2+y^2\biggr)\right] {Y_l^m}^*(\widehat{y-x}).
\end{equation}

In our computation of the effective-range correction to the two-point function of the charge-three operators we will also need to evaluate the following three point functions which involve a descendant of $O_2$, 

\begin{equation}
       \langle 0|T\psi_\uparrow(t,\x)\left( i\partial_{t} + \frac{\nabla^2}{4} \right)O_2(t,\vec{y}){O^{l, m ,s\dagger}_3}(0,\vec 0)|0\rangle, \quad
   \langle 0|T O^{l, m ,s}_3(0,\vec{0}) \left( -i\partial_{t} + \frac{\nabla^2}{4} \right)O^\dagger_2(t,\vec{y})\psi_\uparrow^\dagger(t,\x)|0\rangle .
\end{equation}

Using the fact that $\psi$ and $\psi^\+$ satisfies the free equation of motion when acting on the vacuum from the left and the right, respectively,
\begin{equation}
   \langle 0 | \left( i\d_t + \frac{\nabla_x^2}2 \right)\psi = 0, \qquad
    \left( -i\d_t + \frac{\nabla_x^2}2 \right)\psi^\+ |0\rangle = 0,
\end{equation}
we can write, 

\begin{subequations}\label{o3irrel3ptl01}
\begin{align}
    \langle 0|T \psi_\uparrow(t,\x)\left( i\partial_{t} + \frac{\nabla_y^2}{4} \right)O_2(t,\vec{y}){O^{l, m ,s\+}_3}(0,\vec 0)|0\rangle
    &= \left(i\partial_{t} +\frac{\nabla^2_{x}}{2}+\frac{\nabla^2_{y}}{4}\right)\langle 0|T \psi_\uparrow(t,\x) O_2(t,\vec{y}){O^{l, m ,s\+}_3}(0,\vec 0)|0\rangle,\\
    \langle 0|T O^{l, m ,s}_3(0,\vec{0}) \left( -i\partial_{t} + \frac{\nabla_y^2}{4} \right)O^\dagger_2(t,\vec{y})\psi_\uparrow^\dagger(t,\x)|0\rangle &= \left(-i\partial_{t} +\frac{\nabla^2_{x}}{2}+\frac{\nabla^2_{y}}{4}\right) \langle 0|T O^{l, m ,s}_3(0,\vec{0})O^\dagger_2(t,\vec{y})\psi_\uparrow^\dagger(t,\x)|0\rangle,
\end{align}
\end{subequations}
and hence these three point functions are reduced to the ones that have been computed previously.

\section{Conformal perturbation theory for scattering-length and effective-range corrections}\label{cptfsar}

We are now in a position to address the main question in this paper. As stated in the introduction, NRCFT is relevant for the nuclear reactions involving the emission of a few ($N$, where in this paper $N=2$ or 3) neutrons and one other particle near the end point in the energy spectrum of the latter particle. The cross-section for this process is determined by the imaginary part of the correlator of two unnucleus fields, i.e., local operators carrying same quantum numbers as $N$ neutrons in the NRCFT of unitary fermions. This correlator has a scaling behavior dictated by the scaling dimension of the unnuclear operator. This scaling behavior is violated by the deviation of neutron physics from that of a particle with zero-range, infinite-scattering-length interaction.  This deviation is parameterized by the relevant and irrelevant deformation terms that appear in the effective Lagrangian describing the neutron.  There is only one relevant deformation, corresponding to the finite inverse scattering length, and among the infinite number of irrelevant deformations the one with the lowest dimension corresponds to the effective range.  From now on by ``irrelevant'' deformation we will have in mind the effective-range deformation. 

In Sec.~\ref{sec:charge-two} we have computed the relevant and irrelevant deformation to charge-two operator. Recall that because of particle conservation, at an algebric level, the computation involved integrating over three point functions. Our next goal is to evaluate the scattering length and effective range correction to the $O_3$ two point function. As we see below, this involves integrating over three point functions in true spirit of conformal perturbation theory:
\begin{eqnarray}
    G_{O^{l, m ,s}_3}(t,\x)= G^{\rm CFT}_{O^{l, m ,s}_3}(t,\x) + \frac{i}{4 \pi a } G^{\text{rel}}_{ O^{l, m ,s}_3}(t,\x) - \frac{i r_0}{8 \pi} G^{\text{irrel}}_{ O^{l, m ,s}_3}(t,\x),
\end{eqnarray}
where $G^{\rm CFT}_{O_3}(t,\x)$ is the usual two point function and the relevant and irrelevant deformations are given by Eq.~\eqref{deflag} (here we have simplified the expressions as in the charge-two operator case and used Eq.~\eqref{dimero2norm}. We also omit the indices in the operator $O_3$), 
\begin{subequations}
\begin{align}
  G^{\text{rel}}_{ O_3}(t,\x)&= -16\pi^2i \Theta(t)\!\int\limits^t_0\! dt'\! \int\! d^3 \y^{\,\prime}\, \langle 0| O_3(t,\vec{x}) O^\dagger_2(t',\vec{y}^{\,\prime})O_2(t',\vec{y}^{\,\prime})O^{\+}_3(0,\vec 0)|0\rangle,  \\
  G^{\text{irrel}}_{O_3}(t,\x)&= - 16i\pi^2\Theta(t)\!\int\limits^t_0 dt'\! \int\! d^3\y^{\,\prime}\, \langle 0| O_3(t,\vec{x}) O^\dagger_2(t',\vec{y}^{\,\prime})\biggl( i\overset{\leftrightarrow}{\partial}_{t'} + \frac{\overrightarrow{\nabla}_{y'}^2+\overleftarrow{\nabla}_{y'}^2}{4} \biggr)O_2(t',\vec{y}^{\,\prime})O^{\+}_3(0,\vec 0)|0\rangle.
\end{align}
\end{subequations}

In analogy with our treatment of the correlation function of the charge-two operator, we insert the complete set of states between the operators $O_2$ and $O_2^\+$.  Due to number conservation, we need to insert only single-particle states, so we have (we will write the formula for the relevant deformation; the formula for the irrelevant deformation is written completely analogously)
\begin{equation}
  G^{\text{rel}}_{ O_3}(t,\x)
  = -16\pi^2i \Theta(t)\! \int\limits^t_0\! dt'\! \int\! d^3 \x^{\,\prime} \!\int\! d^3\y^{\,\prime}\, \langle 0| O_3(t,\vec{x}) O^\dagger_2(t',\vec{y}^{\,\prime})|t',\x',\uparrow\rangle\langle t',\x',\uparrow| O_2(t',\vec{y}^{\,\prime})O^{\+}_3(0,\vec 0)|0\rangle  ,
\end{equation}
where
\begin{equation}\label{oneparticlestates}
  |t', \x', \sigma\> = \psi^\+_\sigma(t', \x') |0\>, \qquad \sigma = \uparrow, \downarrow .
\end{equation}
Our deformed correlation functions then takes the form of integrals of the product of product of two three point functions
\begin{subequations}
\begin{align}\label{Grel-3pt-square}
  G^{\text{rel}}_{ O_3}(t,\x) =& -16\pi^2i \Theta(t) \!\int\limits^t_0\! dt' \!\int\! d^3 \x^{\,\prime} \!\int\! d^3\y^{\,\prime}\, \langle 0| O_3(t,\vec{x}) O^\dagger_2(t',\vec{y}^{\,\prime})\psi_\uparrow^\dagger(t',\x^{\,\prime})|0\rangle\langle 0|\psi_\uparrow(t',\x^{\,\prime}) O_2(t',\vec{y}^{\,\prime})O_3^\+(0,\vec 0)|0\rangle, \\
  G^{\text{irrel}}_{O_3}(t,\x)
  =& -16\pi^2i \Theta(t) \!\int\limits^t_0\! dt' \!\int\! d^3\x^{\,\prime} \!\int\! d^3\y^{\,\prime}\nonumber\\
  & \left[\langle 0| O_3(t,\vec{x}) O^\dagger_2(t',\vec{y}^{\,\prime})\psi_\uparrow^\dagger(t',\x^{\,\prime})|0\rangle\langle 0|\psi_\uparrow(t',\x^{\,\prime})\biggl( i\partial_{t'} + \frac{\nabla_{y'}^2}{4} \right)O_2(t',\vec{y}^{\,\prime})O^{\+}_3(0,\vec 0)|0\rangle \biggr.\nonumber\\
  & \left. + \langle 0| O_3(t,\vec{x}) \biggl( -i\partial_{t'} + \frac{\nabla_{y'}^2}{4} \biggr)O^\dagger_2(t',\vec{y}^{\,\prime})\psi_\uparrow^\dagger(t',\x^{\,\prime})|0\rangle\langle 0|\psi_\uparrow(t',\x^{\,\prime})O_2(t',\vec{y}^{\,\prime})O^{\+}_3(0,\vec 0)|0\rangle\right].
\end{align}
\end{subequations}
Thus we see that our deformations can be thought of as integrating over two three point functions. We evaluate these integrals systematically in the following subsections.

Before we present the explicit details, we record the expectation from general arguments of 
boost and scale invariance. In NRCFT, two point functions of primary operators are completely fixed by Galilean-boost and scale invariance. Consider a charge-$N_O$ primary operator $O_N(t,\vec{x})$. The action of the generators of Galilean boosts on a primary operator is given by
\begin{equation}
    [K_i,O_N(t,\vec{x})]= (-i t\partial_{i} + x_i N_O) O_N(t,\vec{x}).
\end{equation}
We can use this and scale invariance to constrain the two point function of primary operators as follows \cite{Nishida_2007}
\begin{equation}
    \langle [K_i,\, O_N(t,\vec{x}) O^\dagger_N(0,\vec 0)] \rangle= 0 \implies  \langle O_N(t,\vec{x}) O^\dagger_N(0,\vec 0) \rangle = f(t) \exp\left(-\frac{i N_O x^2}{2 t}\right),  
\end{equation}
where the scale dependence is contained within the function $f(t)\propto t^{-\Delta}$ (using conformal invariance).  The relevant and irrelevant interactions we are considering, 
\begin{subequations}
\begin{align}
    O_{\rm rel}&= \int\limits_0^t\! dt'\!\int\! d^3\vec{x}^{\,\prime}\, O^\dagger_2 (t',\x^{\,\prime})O_2 (t',\x^{\,\prime}), \\
    O_{\rm irrel}&= \int\limits_0^t\! dt'\!\int\! d^3\vec{x}\, O^\dagger_2 (t',\x^{\,\prime})\biggl( i\overset{\leftrightarrow}{\partial}_t + \frac{\overrightarrow{\nabla}^2+\overleftarrow{\nabla}^2}{4} \biggr)O_2 (t',\x^{\,\prime}),
\end{align}
\end{subequations}
commute with the generator of Galilean boosts. For the relevant deformation, a short calculation shows that,  
\begin{equation}
    [K_i,\,  O_{\rm rel}] = -i\! \int\limits^t_0\! dt'\!\int\! d^3\vec{x}\,\, t \partial_i (O^\dagger_2 (t',\x)O_2 (t',\x)) = 0,
\end{equation}
where we have used $N_{O^\dagger_2}=-N_{O_2}$. Thus the two point function deformed by the relevant deformation still obeys the differential equation as the unperturbed one,
\begin{equation}
  \langle [K_i,\, O_N(t,\vec{x}) O_{\text{rel}} O^\dagger_N(0,\vec 0)] \rangle= 0  \implies \langle  O_N(t,\vec{x}) O_{\text{rel}} O^\dagger_N(0,\vec 0) \rangle \sim \frac{ 1 }{t^{\Delta-\frac{1}{2}}}  \exp\left(-\frac{i N_O x^2}{2 t}\right),
\end{equation}
where the power of $t$ in the pre-exponent is determined by dimension. The irrelevant deformation works out similarly (although a bit more tedious),
\begin{equation}
  \langle [K_i, O_N(t,\vec{x}) O_{\text{irre}l} O^\dagger_N(0,0)] \rangle= 0,  \implies \langle  O_N(t,\vec{x}) O_{\text{irrel}} O^\dagger_N(0,0) \rangle \sim \frac{ 1  }{t^{\Delta+\frac{1}{2}}} \exp\left(-\frac{i N_O x^2}{2 t}\right)  ,
\end{equation}
where we have used, 
\begin{equation}
    [K_i,\, \partial_t O]=-2i\partial_i O- it\partial_i \partial_t O+N_Ox_i \partial_t O,\qquad [K_i,\, \partial^2_x O]=4N_O \partial_i O- i t \partial_x^2 \partial_i O+N_O x_i \partial_x^2 O .
\end{equation}
Fourier transforming these deformations, we record the expectation for the imaginary part of the deformed two point function on grounds of scale and galilean boost invariance, 
\begin{equation}
    \text{Im}\, G_{O_N}(\omega,0) \sim \omega^{\Delta-\frac{5}{2}}\left(1+\frac{C_{\rm rel}}{a \sqrt{\omega}} + r_0 C_{\rm irrel} \sqrt{\omega} \right) .
\end{equation}
Note that the result for the deformation corrections in the correlator of the charge-two operator [Eq.~\eqref{o2finalans}] is consistent with this general expectation. 

\subsection{Test case: chemical potential}
\label{testcase}

Before studying the perturbation of the three-body operator, as a warm up exercise and a non-trivial check of our formalism, we determine the corrections to the dimer propagator due to a chemical potential deformation and obtain agreement with the expectation. The chemical potential is a relevant deformation  given by, 
\begin{equation}\label{deflag2}
  \mathcal L' = \mathcal L_\text{CFT} 
  + \mu \sum_{\sigma=\uparrow\downarrow}\ \psi_\sigma^\dagger \psi_\sigma.
\end{equation}

The first order deformation to the charge-two two point function is given by, 
\begin{equation}\label{go2mu3}
    G_{O_2}(t,\x)= G^{\rm CFT}_{O_2}(t,\x) + i \mu G^{\mu}_{ O_2}(t,\x),
\end{equation}
where $G^{\rm CFT}_{O_2}(t,\x)$ is the unperturbed two-point function (which we have evaluated in Eq.~\eqref{gcfto2}) and the chemical potential deformation is given by Eq.~\eqref{deflag2} (here we have simplified the expressions as in the charge-two operator case), 
\begin{equation}
  G^{\mu}_{ O_2}(t,\x) = -i \Theta(t)\sum_{\sigma}\int\limits^t_0\! dt' \!\int\! d^3\vec y^{\,\prime}\, \langle 0| O_2(t,\vec{x}) \psi_\sigma^\dagger(t',\vec{y}^{\,\prime})\psi_\sigma(t',\vec{y}^{\,\prime})O_2^\+(0,\vec 0)|0\rangle .
  \end{equation}
Analogously to the charge-two operator case, we insert a complete set of single particle states [see Eq.~\eqref{oneparticlestates}] to evaluate the deformed two-point functions. 
\begin{align}\label{gmuo2}
  G^{\mu}_{ O_2}(t, \x)
  &=-i \Theta(t)\sum_{\sigma,\sigma'} \int\limits ^t_0\! dt' \!\int\! d^3 \x^{\,\prime} \!\!\int\! d^3\y^{\,\prime}\, \langle 0|O_2(t,\vec{x}) \psi_\sigma^\dagger(t',\vec{y}^{\,\prime})|t',\x^{\,\prime},\sigma'\rangle\langle t',\x^{\,\prime},\sigma'| \psi_\sigma(t',\vec{y}^{\,\prime})O_2^\+(0,\vec 0)|0\rangle   \nonumber\\
  &= -2i \Theta(t)\int\limits ^t_0\! dt' \!\int\! d^3 \x^{\,\prime} \!\!\int\! d^3\y^{\,\prime}\, \langle 0| O_2(t,\vec{x}) \psi_\downarrow^\dagger(t',\vec{y}^{\,\prime})\psi_\uparrow^\dagger(t',\x^{\,\prime})|0\rangle\langle 0|\psi_\uparrow(t',\x^{\,\prime}) \psi_\downarrow(t',\vec{y}^{\,\prime})O_2(0,\vec 0)|0\rangle .
\end{align}
The three point functions that appear under the integral can be understood as time-ordered correlators. Thanks to our computation in Eq.~\eqref{gmuo21}, we now have all the ingredients to evaluate Eq.~\eqref{gmuo2}. 
We have
\begin{equation}
  i \mu G^{\mu}_{ O_2}(t,\x)
= \frac{\mu}{8\pi^6} \Theta(t)\! \int\limits ^t_0\! dt' \!\int\! d^3 \x^{\,\prime} \!\!\int\! d^3\y^{\,\prime}\,  \frac{1}{t^{\prime2} (t-t')^2|\vec x^{\,\prime}-\y^{\,\prime}|^2} \exp\!\left[\frac{i(\frac{\xp+\yp}{2})^2}{t'}+\frac{i(\xp-\yp)^2t}{4t'(t-t')}+\frac{i(\x-\frac{\xp+\yp}{2})^2}{t-t'}\right].
\end{equation}
In order to evaluate this integral, one can  make the change of coordinates to 
$\x_+=\frac12(\xp+\yp)$, $\x_-= \xp-\yp$.  One then finds
\begin{equation}
  i \mu G^{\mu}_{ O_2}(t,\x) = 
-\frac{\mu \Theta(t)}{2\pi^3 t} \exp\left(i \frac{x^2}{t}\right).
\end{equation}
Putting everything together we get \eqref{go2mu3} to be, 
\begin{equation}
     G_{O_2}(t,\vec{x})=  i\frac{\Theta(t)}{4 \pi^3 t^2} \exp\left(i \frac{x^2}{t}\right) -\frac{\mu \Theta(t)}{2\pi^3 t} \exp\left(i \frac{x^2}{t}\right).
\end{equation}
The Fourier transform is given by [using Eq.~\eqref{ft}], 
\begin{equation}
    G_{O_2}(\omega,\vec{p}) =  -\frac{1}{4\pi \sqrt{\frac{p^2}{4}- \omega}} -\frac{\mu}{4\pi(\frac{p^2}{4}- \omega)^{\frac{3}{2}}}  \,.
\end{equation}
This result can be easily checked independently. The dimer propagator under chemical potential deformation can be computed exactly by noting that the effect of chemical potential is to shift the energy of a charge $N$ operator to $\omega \rightarrow \omega+ N \mu$. This exact result can be obtained diagramatically 
\begin{equation}
    G^{\text{exact}}_{O_2}(\omega,\vec{p}) =  -\frac{1}{4\pi \sqrt{\frac{p^2}{4}- \omega- 2\mu}}= -\frac{1}{4\pi \sqrt{\frac{p^2}{4}- \omega}} -\frac{\mu}{4\pi(\frac{p^2}{4}- \omega)^{\frac{3}{2}}} + O(\mu^2).
\end{equation}
We obtain agreement using two different approaches and this serves as a non trivial test of our conformal perturbation theory techniques.

\subsection{Scattering length deformation}
In this section we evaluate the scattering length correction to the three-body operator two point function for $l=0$ and $l=1$. Let us recall that the relevant deformation is given by Eq.~(\ref{Grel-3pt-square}).  The three-point correlators can be read from Eq.~\eqref{spinlo3o2psi}. Making a change of coordinates to, 
$\x_+=\frac13(\xp+2\yp)$, $\x_-=\yp-\xp$
and using the integrals in the Appendix \ref{usefulint}, one obtains 
\begin{eqnarray}\label{o3reldefgenl}
 G^{\text{rel}}_{ O^{l, m ,s}_3}(t,\x)
     &=&-\frac{64 i \sqrt{2} \pi ^{9/2} |N^{l}_{3pt}|^2 3^s  \Gamma \left(s+\frac{1}{2}\right) \Theta (t)}{(6 l+3)(i t)^{s+2}} \exp\left(\frac{3 i x^2}{2 t}\right) \!\int\limits_0^t\! dt'\, \frac{1}{{t'(t'-t)}^\frac{1}{2}}\nonumber\\
     &=&-2\left(\frac{2}{\sqrt{3}}\right)^6 \frac{27 \pi ^{5/2} 2^{\frac{3}{2}-s} |\phi^l_s(0)|^2  \Gamma \left(s+\frac{1}{2}\right) \Theta (t) }{(2 l+1)  \Gamma (s+1)^2(i t)^{s+2}}\exp\!\left(\frac{3 i x^2}{2 t}\right),
\end{eqnarray}
where we have used the normalization of the spherical harmonics noted in Eq.~\eqref{sphharmnorm} and the generalization of \eqref{eq:Nls_defn} to any $l$. Combining Eqs.~\eqref{o32ptgenl} and \eqref{o3reldefgenl},  the first order relevant deformation gives us the following two-point function 
\begin{equation}
 G_{O^{l, m ,s}_3}(t,\x)
 = -i \left[ \frac{ 2^{-s+\frac{3}{2}}f^l_s}{\Gamma(1+s)\pi^{\frac{3}{2}}(i t)^{s+\frac{5}{2}}}  - \frac{\pi ^{5/2} 2^{\frac{13}{2}-s} |\phi^l_s(0)|^2  \Gamma \left(s+\frac{1}{2}\right) }{ \pi a (2 l+1)  \Gamma (s+1)^2(i t)^{s+2}} \right] \Theta(t)\exp\!\left(\frac{3 i x^2}{2 t}\right).
 \end{equation}
Using the Fourier transform \eqref{ft} we obtain
\begin{equation}
  G_{O^{l, m ,s}_3}(\omega, \vec{p})=
  \frac{2^{3-s}\pi}{3^{\frac32}\Gamma^2(s+1)}\left[\frac{f^l_s}{\sin(\pi s)}\left( \frac{p^2}{6}- \omega\right)^{s}-\frac{32\pi^3 |\phi^l_s(0)|^2}{(2l+1)\cos(\pi s)}\frac 1a \left( \frac{p^2}{6}- \omega\right)^{s-\frac12}\right].
\end{equation}
The imaginary part of the correlator is is given by, 
\begin{equation}
   \text{Im}~G_{O^{l, m ,s}_3}(\omega, \vec{p})
    = -N^{l,s}_0\Theta\!\left(\omega-\frac{p^2}{6}\right) \left(\omega-\frac{p^2}{6}\right)^{s} \left( 1 + \frac{C^{l,s}}{a\sqrt{\omega-\frac{p^2}6}} \right),
\end{equation}
where 
\begin{equation}\label{Nijdef2}
    N^{l,s}_0 = \frac{2^{3-s}\pi f_s^l}{3^\frac32\Gamma^2(s+1)}\, ,\qquad C^{l,s}= \frac{32\pi^3|\phi^l_s(0)|^2}{(2l+1)f^l_s} \,.
\end{equation}
We evaluate the innerproduct $f^l_s$ defined in \eqref{normfs} for $l=0$ and $l=1$. We present the numerical values of $N^{l, s}_0$ and $C^{l, s}$ in the table \ref{table3} for $l=0$ and $l=1$, relegating the details to Appendix \ref{fscalc}.  We also note that for the $n$th $l=0$ operator at large $n$, from Eq.~(\ref{sn-asympt}) we find $C^{l=0,s}=2+O(n^{-1})$.  The trend of $C^{l=0,s}$ converging to 2 can already be seen in Table \ref{table3}.
\begin{table}
	\begin{center}
		\begin{tabular}{|c|c|c|c|}
			\hline
			$s$ & $N_0^{l=0,s}$ & $C^{l=0,s}$ \\
			\hline
			$~2.16622~$ & 3.12650 & $~4.14759~$ \\
			\hline
			 5.12735 & $3.43747\times 10^{-3}$ &  1.72021 \\
			\hline
			 7.11448 & $3.83887 \times 10^{-7}$  &  2.16335 \\
			\hline
          8.83225 & $~7.81789\times 10^{-11}$ &  2.02332 \\
          \hline
		\end{tabular}
  \qquad\qquad
  \begin{tabular}{|c|c|c|c|}
			\hline
			$s$ & $N_0^{l=1,s}$ &  $C^{l=1,s}$ \\
			\hline
			$~1.77272~$ & 182.046 &  $~2.64179~$\\
			\hline
			 4.35825 & 0.294377&  2.06642\\
           \hline
			 5.71643 & $~1.98465 \times 10^{-3}\,$ &  2.24789\\
			\hline
			 8.05319 & $~1.02679\times 10^{-7}\,$ &  1.84982\\
			\hline
		\end{tabular}
	\end{center}
\caption{Numerical values of $N^{l,s}_0$ and $C^{l,s}$ [defined in Eq.~\eqref{Nijdef2}] for $l=0,1$ and a few lowest values of $s$.}
\label{table3}
\end{table}

\subsection{Effective-range correction}\label{effrangecorr}
We now proceed to evaluate the effective-range correction. The correction to the two-point function of $O_3$ is
\begin{align}
    G^{\text{irrel}}_{ O_3}(t,\x)
  =& -16\pi^2i \Theta(t)\!\int\limits^\infty_{-\infty}\! dt' \!\int\! d^3\xp \!\!\int\! d^3\yp\nonumber\\
  & \left[\langle 0| O^{l=0 ,s}_3(t,\vec{x}) O^\dagger_2(t',\yp)\psi_\uparrow^\dagger(t',\xp)|0\rangle
  \langle 0|\psi_\uparrow(t',\xp)\biggl( i\partial_{t'} + \frac{\nabla_{y'}^2}{4} \biggr)O_2(t',\yp)O^{l=0 ,s}_3(0, \vec 0)|0\rangle \right.\nonumber\\
  & \left. + \langle 0| O^{l=0 ,s}_3(t,\vec{x}) \biggl( -i\partial_{t'} + \frac{\nabla_{y'}^2}{4} \biggr)O^\dagger_2(t',\yp)\psi_\uparrow^\dagger(t',\xp)|0\rangle\langle 0|\psi_\uparrow(t',\xp)O_2(t',\yp)O^{l=0 ,s}_3(0,\vec 0)|0\rangle\right].
\end{align}
We use the three-point functions listed in Eqs.~\eqref{o3irrel3ptl01} 
and perform the integrals to get, 
\begin{align}
G^{\text{irrel}}_{ O_3}(t,\x)=&-16\pi^2|N^{l=0}_{3pt}|^2i\Theta(t) \int\limits^{\infty}_{-\infty} dt' \left[  -\frac{\Theta(t-t')\Theta(t')}{t'^{3/2} (t-t')^{3/2}} \frac{st(s-1)}{(2s-1)}+\frac{\Theta(t')}{\sqrt{t'}}\partial_{t'}\frac{\Theta (t-t')}{\sqrt{t-t'}} -\frac{\Theta(t-t')}{\sqrt{t-t'}}\partial_{t'}\frac{\Theta (t')}{\sqrt{t'}}  \right]\nonumber\\
  &\times 4 \sqrt{2}\pi ^{5/2} 3^{s-1} t^{-s-2} \Gamma \left(s+\frac{1}{2}\right) \exp\!\left[-\frac{i \left(\pi  s t-3 x^2\right)}{2 t}\right].
  \end{align}

The integral is divergent, and the divergent piece can be interpreted as wave-function renormalization of $O_3$.  We are interested in the finite piece, which can be read out from the integrals listed in the appendix [see Eqs.~\eqref{irrelint1} and \eqref{irrelint2}].  We conclude that the first order deformation has no finite piece.  This is due to, in particular, the presence of $\Gamma(1-\frac32-\frac32)=\infty$ in the denominator.  
A smiliar analysis can be done for $l=1$ and we find that the conclusion remains the same. 

The absence of the finite piece can be understood as a feature of integer dimensions. In Appendix \ref{gendimd} we show that indeed for non-integer dimensions, we do have a finite piece along with the wave-function renormalization. It might be a worthwhile exercise to try and understand vanishing of the irrelevant deformation in this NRCFT from the perspective of correlation function behaviour of a higher dimensional relativistic CFT where momentum space renormalisation schemes have been studied in detail and explicit counterterms classified \cite{Guica:2010sw, Bzowski:2015pba}\footnote{We thank Kostas Skenderis for suggesting this possibility.}. In a similar manner, in Appendix \ref{sec:irrel-free} we observe the absence, near the free fixed point, of the first-order correction due to the scattering length (which is an irrelevant deformation near that fixed point).

As a possibly relevant anecdote, let us consider the first order irrelevant deformations of two point functions of relativistic conformal field theories \cite{Sen:2017gfr, Berenstein:2016avf}\footnote{We thank Bruno Balthazar and Yuji Tachikawa for discussions on this.}. 
Consider the deformation of two point function of primary operators $O$ by an irrelevant operator $\Tilde{O}$ of dimension $\Delta$. The Lagrangian of the system is given by 
\begin{equation}
    \mathcal L= \mathcal L_\text{CFT} + \lambda \!\int\! d^dx\, \lambda \Tilde{O} .
\end{equation}
The first order deformation of the scalar primary operator two point function is given by \cite{Sen:2017gfr},
\begin{eqnarray}
    \langle O(x_1)O(0) \rangle_\lambda \sim \frac{\Gamma (h) \Gamma (\Delta -h) \Gamma \left(\frac{1}{2} (2 h-\Delta )\right)^2}{x_1^{\frac{\Delta-2h}{2}}\Gamma \left(\frac{\Delta }{2}\right)^2 \Gamma (2 h-\Delta )},\qquad h=\frac{d}{2}\,.
\end{eqnarray}
For $\Delta=d+n$, this expression vanishes when $n$ is odd, it is finite when $n$ is non-integer with no divergent pieces for either case. It has a finite as well as a divergent piece when $n$ is even integer. We seem to be witnessing a similar phenomenon for NRCFT irrelevant deformations. It would be interesting to justify these observations from symmetry reasons for both relativistic as well as non-relativistic CFTs.

\section{Conclusion}
\label{sec:conclusion}

In this paper we have studied the notion of local operators in the NRCFT describing fermions at unitarity.  Because the latter theory is basically nonrelativistic quantum mechanics, these local operators can be defined explicitly through formulas expressing the matrix elements of these operators between two arbitrary states with the wave functions of these states.  In principle, this allows one to compute any correlation function in the NRCFT by inserting the complete sets of states between the operators; the ``only'' requirement is that one needs to know the general solution of the Schr\"odinger equation for the relevant number of particles that appear in the complete sets.  In practice, for unitarity fermion, exact analytic solutions to the Sch\"odinger equation are known for one, two and three particles, which already allows the solution of a number of nontrivial problems, including the computation of corrections to the unnuclear scaling law proportional to the inverse scattering length and effective range.  

We have found that the first effective range ($r_0$) correction to the imaginary part of the two-point functions vanishes in three spatial dimensions.  This means that effective-range corrections appear first at order $r_0^2$.  At this order, however, the nuclear reaction cross sections receive a contribution from the two-point correlator of the descendant $\vec\nabla O_3$, and it is not clear if the $r_0^2$ correction can be cleanly separated from the contribution from that descendant. 

There are several avenues to be explored. It would be worthwhile to try to sum up the relevant perturbations to the two-point function of the charge-three operator at the unitary fixed point to all orders in $1/a$ and derive an expression that gives the two-point function along the renormalization group (RG) trajectory connecting the interacting fixed point and the free theory.  For the charge-two operator $O_2$, the exact expression is known and reproduces the Watson-Migdal formula for the emission of two neutrons~\cite{Watson:1952ji,Migdal:1955}.  The exact three-point function of $O_N$ along the RG flow will gives the ``unnuclear'' behavior of processes with the production of $N$ neutron from the scale $\hbar^2/m_nr_0^2$  down to arbitrarily small energy.  Note that although the leading $r_0$ correction vanishes near the unitarity fixed point, it does not have to vanish along the whole RG flow.

In Appendix \ref{gendimd}, we have worked out the quantum numbers of the charge-three operator in general dimensions ($2<d<4$). It would be satisfying to evaluate and crosscheck our data  for the charge-three operator, obtained in this work, using  $\epsilon$ expansion perturbation theory for unitary fermions around $d=2$ and $d=4$ \cite{Nishida_2007}.

In Refs.~\cite{Son:2008ye, Balasubramanian:2008dm}, a toy model for holographic construction of Schr\"odinger invariant quantum field theories was proposed. Although these are not realistic duals of unitary fermions, these constructions provide a way to explore universal features of NRCFTs through dimensional reduction of usual relativistic quantum field theories. More precisely, in Ref.~\cite{Fuertes:2009ex}, the authors showed that light cone reduction of tree level AdS three point correlators is in agreement with $\langle O_2 \psi^\dagger \psi^\dagger\rangle$ three point correlator upto a normalisation constant. It would be interesting to extend this geometric picture to correlators of the charge-three operators evaluated in this work. One might even hope that such an exercise sheds light on the vanishing of the first order irrelevant perturbation as well. 

A natural question is to extend our analysis of defining local operators and their correlation functions, to higher charges.  A knowledge of the behavior of the two-point correlator of the charge-four operator may be useful for interpreting the recent observation of nontrivial correlation in the four-neutron system~\cite{Marques:2001wh,Kisamori:2016jie,Duer:2022ehf}.  One may expect that the present method outlined in this paper is not the most efficient way of going about this since solving the Schr\"odinger equation for many-body wave function with short range interaction is a difficult task. However, recent years has seen a lot of progress in understanding non-relativistc quantum field theories in the large charge regime, where the inverse charge acts as a perturbative parameter for stongly coupled theories \cite{Kravec:2018qnu, Favrod:2018xov, Kravec:2019djc}. The large charge sector is described by a Nambu-Goldstone EFT, associated with the $U(1)$ particle number \cite{Son:2005rv} and properties of large charge operators have been studied in detail \cite{Hellerman:2020eff, Pellizzani:2021hzx, Hellerman:2021qzz}. It would be interesting to compute correlation functions of the charge $Q$ operators,  $\langle O_Q O_Q^\dagger\rangle$ and $\langle O_{Q+q}  O^\dagger_q O_Q^\dagger\rangle$  for $Q\gg q$ in this formalism. Using the non-relativistic state-operator correspondence, one may expect such correlation functions to be respectively evaluated as inner product or matrix elements of $O_q$ in eigenstates of the harmonic potential, which can be evaluated semiclassically using the Nambu-Goldstone EFT.\footnote{The relativistic counterpart of this exercise was carried out in Refs.~\cite{Monin:2016jmo, Cuomo:2020rgt}.}    

Another direction of exploration is bosonic systems with unitarity interaction. Such systems which is approximately realized in nature by the $^4$He atoms, as well as by the $\alpha$-particles in nuclear physics.  A treatment of the Efimov effect, as well as of the Coulomb interaction in the case of the $\alpha$-particles, would be essential.  Neutral charm mesons also provide a playground for NRCFT~\cite{Braaten:2021iot}. We leave these threads for future work. 
\acknowledgements

This work is supported, in part, by the NSF Grant No.\ PHY2014195, U.S.\ DOE Grant No.\ DE-FG02-13ER41958, Chandrasekhar fellowship at the University of Chicago, Kadanoff fellowship at the University of Chicago and by the Simons Collaboration on Ultra-Quantum Matter, which is a grant from the Simons Foundation (651440, D.T.S.).  The authors thank Bruno Balthazar, Luca Delacr\'etaz,  Hans-Werner Hammer, Simeon Hellerman, Xiavier Layronas, Kostas Skenderis, Yuji Tachikawa, and F\'elix Werner for discussions. SDC thanks Mainz Institute for Theoretical Physics (MITP) of the Cluster of Excellence $\text{PRISMA}^+$ (Project ID 39083149), Organisers of the workshop ``Thermalisation in Conformal Field Theories" and EPFL for hospitality and opportunity to present the work. RM would like to thank, ICTS, TIFR, IISc and organisers of the ``Paths to QFT 2023" workshop for hospitality, stimulating discussions and opportunity to present this work. 

\appendix

\section{Useful integrals}\label{usefulint}
In this Appendix we list some useful integrals we use in the main text,
\begin{subequations}
\begin{eqnarray}
    \int\! d^3 x\, e^{i \vec{x}^2 \alpha + i \vec{x}\cdot \vec{\beta}+I \gamma} &=&\left(\frac{\sqrt{\pi }}{\sqrt{-i \alpha }}\right)^3 \exp \biggl( i \gamma -\frac{i {\beta} ^2}{4 \alpha }\biggr), \\
\int\limits_0^\infty\! dk\, \frac{e^{i \alpha  k^2} \cos (\beta  k)}{\beta } &=& \frac{\sqrt{\pi } }{2 \sqrt{-i \alpha } \beta } \exp\biggl(-\frac{i \beta ^2}{4 \alpha }\biggr)\,,\\
\int\limits_0^\infty\! dk\, k^{s+1} e^{i \alpha  k^2} J_s(k \gamma )&=&2^{-s-1}  (-i \alpha )^{-s-1} \gamma ^s \exp\biggl(-\frac{i \gamma ^2}{4 \alpha }\biggr) .    
\end{eqnarray}
\end{subequations}

We outline the method of doing divergent integrals encountered in Sec.~\ref{effrangecorr} in order to extract the finite piece. 
\begin{equation}
    A(t) = \int\limits^\infty_{-\infty}\! dt'\, \frac{\Theta(t-t')\Theta(t')}{(t-t')^\alpha t'^\beta} \,.
\end{equation}
We first evaluate the Fourier transform $A(\omega)$ via analytic continuation (Since the Fourier transform is valid only for $\alpha, \beta <1$) and inverse Fourier transform to get the position-space answer. 
\begin{eqnarray}
    A(\omega)= \Gamma(1-\alpha)\Gamma(1-\beta) (-i \omega)^{\alpha+\beta-2} .
\end{eqnarray}
The inverse fourier transform gives us, 
\begin{eqnarray}\label{irrelint1}
    A(t)= \frac{\Theta(t)\Gamma(1-\alpha)\Gamma(1-\beta)}{\Gamma(2-\alpha-\beta)} t^{-\alpha-\beta+1}\,.
\end{eqnarray}
Similarly, we record the following integrals that we use, 
\begin{equation}\label{irrelint2}
    \int\limits^\infty_{-\infty}\! dt'\, \partial_{t'}\!\!\left[\frac{\Theta(t-t')}{(t-t')^\alpha }\right]\frac{\Theta(t')}{t^{\prime\beta} } = - \!\int\limits^\infty_{-\infty}\! dt'\, \frac{\Theta(t-t')}{(t-t')^\alpha }\,\partial_{t'}\!\!\left[\frac{\Theta(t')}{(t')^\beta }\right] = -\frac{\Theta(t)\Gamma(1-\alpha)\Gamma(1-\beta)}{\Gamma(1-\alpha-\beta)} t^{-\alpha-\beta},
\end{equation}

\section{Operator product expansion}\label{opeappendix}
In this Appendix we present an expression for operator product expansion of charge-two and charge-one primary operators at equal times. Schematically the OPE\footnote{See Refs.~\cite{Golkar:2014mwa,Goldberger:2014hca} and references therein for some progress in understanding OPE structure of NRCFTs in general.} can be stated as 
\begin{equation}
    O_2(t,\y) \psi_\uparrow(t,\x) = \sum_i f_i \left(t,\x-\y, \frac{2\y+\x}{3}\right) O_i\left(t,\frac{2\y+\x}{3}\right),
\end{equation}
where $f_i$ is a functional of its arguments and encodes the contribution of descendants of the primary operator $O_i$, the label $i$ denotes the different primary operators that appear in the OPE. We provide a closed form expression for the functional $f_i$ in this Appendix for the particular class of primaries $O^{l,m,s}_3$. We take our three point function \eqref{spinlo3o2psi},    
\begin{align}\label{spinlo3o2psiB}
      \langle 0| O^{l,m,s}_3(t,\vec{x}) O^\dagger_2(t',\yp)\psi_\uparrow^\dagger(t',\xp)|0\rangle &= N^{l}_{3pt} \frac{|\yp-\xp|^{s-1}}{(t-t')^{s+\frac52}}{Y_l^m}^*(\widehat{\yp-\xp} ) \exp\!{\left[\frac{i3(\vec{x}-\frac{\x'+2\y'}{3})^2}{2(t-t')}+\frac{i(\xp-\yp)^2}{3(t-t')}\right]} ,\nonumber\\
      N^{l}_{3pt} &= - \frac{i^{-s-\frac{5}{2}} 2^{-\frac{s}{2}+1} 3^{\frac{1-s}{2}} {\phi^*}_s^{l}(0)}{\pi ^{3/2} \Gamma (s+1)} \,.
\end{align}
In the OPE limit, this can be schematically written as 
\begin{eqnarray}\label{opeo2psi}
   \lim_{\x' \rightarrow \y' } \langle 0| O^{l,m,s}_3(t,\vec{x}) O^\dagger_2(t',\yp)\psi_\uparrow^\dagger(t',\xp)|0\rangle &=&f_{l, m, s}\left(t',\xp-\yp, \frac{2\yp+\xp}{3}\right)\left\langle O^{l,m,s}_3(t,\vec{x}) {O^{l,m,s\dagger}_3}\left(t',\frac{2\yp+\xp}{3}\right)\right\rangle \nonumber\\
   &=&N^l_{2pt}f_{l, m, s}\left(t',\xp-\yp, \frac{2\yp+\xp}{3}\right) \frac{\exp\!{\left[\frac{3 i (\vec{x}-\frac{2\y'+\x'}{3})^2}{2(t-t')}\right]}}{( t-t')^{s+\frac{5}{2}}}  ,
\end{eqnarray}
where $N^l_{2pt}=\frac{3^{\frac{3}{2}}2^{-s-\frac{3}{2}}f^l_s}{\Gamma(1+s)\pi^{\frac{3}{2}}(i )^{s+\frac{5}{2}}}$. Comparing Eqs.~\eqref{spinlo3o2psiB} and \eqref{opeo2psi}, we can now express $f_{l, m, s}(t',\x'-\y', \frac{2\y'+\x'}{3})$ as a differential operator. More precisely, 
\begin{eqnarray}
   N^l_{2pt}f_{l, m, s}\left(t',\xp-\yp, \frac{2\yp+\xp}{3}\right) \frac{\exp{\left(\frac{3 i (\vec{x}-\frac{2\y'+\x'}{3})^2}{2(t-t')}\right)}}{( t-t')^{s+\frac{5}{2}}} &=& N^{l}_{3pt}\left( \frac{|\yp-\xp|^{s-1} {Y_l^m}^*(\widehat{\yp-\xp} )}{(t-t')^{s+\frac52}}\right)\nonumber\\
   &&\times \exp{\left(\frac{i3(\vec{x}-\frac{\x'+2\y'}{3})^2}{2(t-t')}+\frac{i(\xp-\yp)^2}{3(t-t')}\right)} .
   \end{eqnarray}
After a bit of algebra, we can write, 
\begin{eqnarray}
    f_{l, m, s}(t,\x_-, \x_+) = \frac{N^l_{3pt}}{N^l_{2pt}} |\x_-|^{s-1} {Y_l^m}^*(\hat{\x}_- ) \, _0F_1\!\left(;s+1;- \frac{|\x_-|^2}{3}\biggl(\frac{\partial^2_{\x_+}}{6}-i\partial_t\biggr)\right),
\end{eqnarray}
where $\, _0F_1(;b;z)$ is a hypergeometric function which has the formal series expansion 
\begin{eqnarray}
    \, _0F_1(;b;z)=\sum_{k=0}^\infty \frac{z^k}{k! (b)_k}\,.
\end{eqnarray}

\section{Computing the overlap $f^l_s$}\label{fscalc}
We illustrate the numerical integration of \eqref{normfs}, performed in the main text.
For this purpose let us recall
\begin{equation}\label{fellapp}
    f^l_s=\int d\bar{\Omega} \left|(1-P_{13})^2\frac{\phi^l_s(\alpha)}{\sin{2\alpha}}Y_l^m(\hat{\rho})\right|^2.
\end{equation} 

Let us explicitly see the action of the operator $P_{13}$ on the coordinates $\alpha$ and $\hat{\rho}$.
\begin{eqnarray}
   \alpha&=&\arctan\frac{|\vec{r}|}{|\vec{\rho}|}\,, \qquad \hat{\rho} , \nonumber\\
    \text{where}~~~
    \vec{r}&=& \vec{x}_2-\vec{x}_1, \qquad \vec{\rho}= \frac{2}{\sqrt{3}}\left( \vec{x}_3- \frac{\vec{x}_1+\vec{x}_2}{2} \right).
    \end{eqnarray}
This implies, 
\begin{eqnarray}\label{p13action}
    P_{13}(\Vec{r}) &=& \frac{\Vec{r}-\sqrt{3}\,\vec{\rho}}{2}, \quad P_{13}(\Vec{\rho}) = -\frac{\sqrt{3}\,\Vec{r}+\vec{\rho}}{2}, \quad
    P_{13}(\alpha) = \tan^{-1}\sqrt{\frac{1+2\cos^2 \alpha- \sqrt{3} \sin 2\alpha\, \hat{r}\cdot \hat{\rho}}{2\sin^2 \alpha+1+ \sqrt{3} \sin 2\alpha\, \hat{r}\cdot \hat{\rho}}}\,,\nonumber\\
    P_{13}(\hat{\rho}) &=&  -\frac{\sqrt{3}\, \vec r + \vec \rho}{|\sqrt{3}\, \vec r + \vec \rho|}=-\frac{\left(\sqrt{3}\sin{\alpha}\,\hat{r}+\cos{\alpha}\,\hat{\rho}\right)}{\sqrt{2\sin^2 \alpha+1+ \sqrt{3} \sin 2\alpha\, \hat{r}\cdot \hat{\rho}}} \,,\nonumber\\
    \text{where}\nonumber\\
    \hat{r}&=&(\sin \theta_r \sin \phi_r, \sin \theta_r \cos \phi_r, \cos \theta_r),\qquad \hat{\rho}=(\sin \theta_\rho \sin \phi_\rho, \sin \theta_\rho \cos \phi_\rho, \cos \theta_\rho).
\end{eqnarray}

For $l=0$ therefore the overlap integral is 
\begin{equation}\label{fs0}
    f^0_s =  4\pi\! \int\limits^{\pi}_0\! \sin \theta_r d\theta_r \!\int\limits^{\pi}_0\! \sin \theta_\rho d\theta_\rho \!\int\limits_{-\pi}^\pi\! d\phi_- \! \int\limits_0^{{\pi}/{2}}\! \sin^2 2\alpha
    \left[\frac{2|\phi^0_s(\alpha)|^2}{\sin^2 2\alpha} - \frac{\phi^0_s(\alpha){\phi^0_s}^*(P_{13}(\alpha))+\phi^0_s(P_{13}(\alpha)){\phi^0_s}^*(\alpha)}{\sin 2\alpha~ \sin 2P_{13}(\alpha)} \right],
\end{equation}
where we have used $\phi_-= \frac12(\phi_r-\phi_\rho)$, $\phi_+= \frac12(\phi_r+\phi_\rho)$.\footnote{The measure changes as $$\int^{2\pi}_{0} d\phi_{r}\int^{2\pi}_{0}d\phi_{\rho}=\int_{-\pi}^{\pi} d\phi_- \int_{\phi_-}^{2\pi+\phi_-} d\phi_+ $$. Hence we have for functions with cylindrical symmetry, 
\begin{eqnarray}
    &&2\int^{\pi}_0~ \sin \theta_r d\theta_r \int^{\pi}_0~ \sin \theta_\rho d\theta_\rho \int_{0}^{2\pi}~ d\phi_r~\int_{0}^{2\pi}~ d\phi_\rho \int_0^{\frac{\pi}{2}}d\alpha~ \sin^2 2\alpha~ g(\phi_-,\theta_\rho, \theta_r,\alpha) \nonumber\\
    &&= 2(2\pi) \int^{\pi}_0~ \sin \theta_r d\theta_r \int^{\pi}_0~ \sin \theta_\rho d\theta_\rho \int_{-\pi}^\pi~ d\phi_-\int_0^{\frac{\pi}{2}} d\alpha~ \sin^2 2\alpha ~ g(\phi_-,\theta_\rho, \theta_r,\alpha).
\end{eqnarray} } Similarly for the $l=1$ case,
\begin{eqnarray}\label{fs1}
    f^1_s&=& 4\pi\! \int\limits^{\pi}_0
    \!\sin \theta_r d\theta_r \!\int\limits^{\pi}_0\!\sin \theta_\rho d\theta_\rho \!\int\limits_{-\pi}^\pi\! d\phi_- \!\! \int\limits_0^{{\pi}/{2}}\! d\alpha\, \sin^2 2\alpha \nonumber\\
    &&\left[\frac{2|\phi^1_s(\alpha)|^2\cos^2 \theta_\rho}{\sin^2 2\alpha} -\frac{\left(\phi^1_s(\alpha){\phi^1_s}^*(P_{13}(\alpha))+\phi^1_s(P_{13}(\alpha)){\phi^1_s}^*(\alpha)\right)\cos \theta_\rho P_{13}(\cos \theta_\rho)}{\sin 2\alpha~ \sin 2P_{13}(\alpha)} \right],
\end{eqnarray}
where we use \eqref{p13action} and, 
\begin{equation}
    \hat{r}\cdot \hat{\rho}= \cos \theta_r \cos\theta_\rho + \cos (2\phi_-) \sin \theta_r \sin\theta_\rho,\quad P_{13}(\cos{\theta_{\rho}})=-\frac{\sqrt{3}\sin{\alpha}\cos{\theta_{r}}+\cos{\alpha}\cos{\theta_{\rho}}}{\sqrt{2\sin^2 \alpha+1+ \sqrt{3} \sin 2\alpha\, \hat{r}\cdot \hat{\rho}}}, \quad Y^{l=1}_{m=0}(\hat{\rho})= \cos \theta_\rho .
\end{equation}
We have used  cylindrical symmetry to simplify the inner product $\hat{r}\cdot \hat{\rho}$. Integrals \eqref{fs1} and \eqref{fs0} are now numerically evaluated using the Gauss-Kronrod method. We note that the numerical integral has potential convergence issues at $\alpha={\pi}/{3}$ (these are absent when one tries to do them analytically \cite{werner:tel-00285587}). We isolate an $\epsilon\sim 10^{-10}$ neighbourhood around this region and perform the integral and obtain agreement upto $10^{-6}$ with the analytic result for $l=0$ in Ref.~\cite{werner:tel-00285587}.
\begin{equation}
    f^{l=0}_s =  \frac{32\pi^2}s \sin \left(\frac{\pi  s}{2}\right) \left[\frac{\pi s}{2}  \sin \left(\frac{\pi  s}{2}\right) -\cos\left(\frac{\pi  s}{2}\right) -\frac{2 \pi  }{3 \sqrt{3}} \cos\left( \frac{\pi  s}{6}\right) \right].
\end{equation}
We present the results of numerical integration of \eqref{fellapp} for $l=0,1$ for various values of $s$ in Table \ref{tablea1}.

 \begin{table}
	\begin{center}
		\begin{tabular}{|c|c|}
 \hline
  $s$& $f_s^{l=0}$ \\		
			\hline
			$~2.16622~$ & $~15.9415~$ \\
			\hline
			 5.12735 & 554.016 \\
			\hline
			 7.11448 & 443.969 \\
			\hline
          8.83225 & 457.113 \\
          \hline
		\end{tabular}
  \qquad\qquad
  \begin{tabular}{|c|c|}
  \hline
  $s$& $f_s^{l=1}$ \\
  \hline
       $~1.77272~$ & $~345.377~$ \\
			\hline
			 4.35825 & 2174.79\\
    \hline
			 5.71643 & 3915.43\\
			\hline
			 8.05319 & 11514.6\\
    \hline
  \end{tabular}
  \end{center}
\caption{$f_s^l$ for $l=0,1$.}
\label{tablea1}
 \end{table}

\section{Analysis in general dimensions $2<d<4$}\label{gendimd}
In this appendix we define the charge-two and charge-three operators in general dimensions $2<d<4$. We also compute the irrelevant deformation of their respective two point functions and show that it has a finite piece in non-integer dimensions.
\subsection{$O_2$ in general dimensions}
In $d$ dimensions, the two particle wave function $\Psi_d^{(2)}(\Vec{x}_1,\Vec{x}_2)$ obeys the following Bether-Peierls boundary condition
\begin{equation}\label{bp2d2}
    \lim_{ \Vec{x}_1\rightarrow \Vec{x}_2} \Psi_d^{(2)}(\Vec{x}_1,\Vec{x}_2)\sim \frac{1}{|\Vec{x}_1-\Vec{x}_2|^{d-2}}  .
\end{equation}
As a result, the finite charge-two operator is therefore defined as, 
\begin{equation}\label{O2matrixelgend}
\langle 0| O^d_2(0,\frac{\vec{x_1}+\vec{x_2}}{2}) |\Psi_d^{(2)}\rangle = \lim_{ \Vec{x}_1\rightarrow \Vec{x}_2} |\Vec{x}_1-\Vec{x}_2|^{d-2} \Psi_d^{(2)}(\Vec{x}_1,\Vec{x}_2) .
\end{equation}
where, as before, the factor $|\Vec{x}_1-\Vec{x}_2|^{d-2}$ ensures that the matrix elements are finite despite the singular boundary condition. It is easy to see that the charge-two operator in general dimensions still has $\Delta_{O^d_2} =2$. In $d$ dimensions, the Schr\" odinger wave function for the two particle states with, zero relative angular momentum, becomes, 
\begin{equation}
    -\frac{1}{2} \biggl(\frac{1}{2} \nabla^2_{R_{\text{rm}}} + 2\nabla^2_{r}\biggr) \Psi^{(2)}_d(\vec R_{\text{cm}},r) = E \Psi^{(2)}_d, \qquad E= \frac{ P_{\text{cm}}^2}{4}+ k^2,
\end{equation}
where 
\begin{eqnarray}
    \vec{R}_{\rm cm}&=&\frac{\vec{x}_1+\vec{x}_2}{2},\qquad\vec{r}=\vec{x}_1-\vec{x}_2,\nonumber\\ \nabla^2_{r}&=& \frac{\partial^2}{ \partial r^2} + \frac{d-1}{r}\frac{\partial}{\partial r}\,.
\end{eqnarray}

We follow the same algorithm of quantizing this wave function in a box to get the normalized wave function solution with quantized eigenvalues.  
\begin{equation}
    \Psi_d^{(2)}(\vec{R}_{\rm cm}, r)=\frac{e^{i \vec{P}_{cm} \cdot \vec{R}_{\rm cm}}}{\sqrt{V R}}\left(\frac{\pi}{2}\right)^{\frac{1}{2}-\frac{d}{4}}\sqrt{\Gamma\left(\frac{d}{2}\right)} \sqrt{k} r^{1-\frac{d}{2}} Y_{\frac{d-2}{2}}(k r), \qquad k=\frac{(2n +1)\pi}{2 R},\quad n \in \mathbb{Z} .
\end{equation}
The undeformed two point function takes the form 
\begin{equation}
   \langle 0| O^d_2(t,\vec{x}) O^{d\dagger}_2(0,\vec 0)|0\rangle = \frac{2^{d-5} (d-2) \pi ^{-d-1} \csc \left(\frac{\pi  d}{2}\right) \Gamma \left(\frac{d}{2}-1\right)^2 }{t^2} \exp{\left(\frac{i x^2}{t}\right)}\,.
\end{equation}

We illustrate that the first order irrelevant deformation of the charge-two operator two point function remains finite for $d\neq 3$ and the divergences cancel 
\begin{eqnarray}
    G^{\text{irrel}}_{ O^d_2}(t,\vec{x})
     &=&- 16i\pi^2\Theta(t)\!\int\limits^t_0\! dt'\! \int d^d\yp 
 \bigg[\langle 0| O^d_2(t,\vec{x}) {O^{d\dagger}_2}(t',\yp)|0\rangle \langle 0|\biggl( i\partial_{t'} + \frac{\nabla^2_{y'}}{4} \biggr)O^d_2(t',\yp){O^{d\+}_2}(0,\vec{0})|0\rangle \nonumber\\
    && \qquad\qquad \qquad\qquad\qquad
    +\langle 0| O^d_2(t,\vec{x}) \biggl( -i\partial_{t'} + \frac{\nabla^2_{y'}}{4} \biggr){O^{d\+}_2}(t',\yp)|0\rangle \langle 0|O^d_2(t',\yp){O^d_2}^\+(0,\vec{0})|0\rangle\biggr]\nonumber\\
    &=& N^d_{O_2} \frac{i \pi ^{\frac{d}{2}+1} t^{\frac{d}{2}-4} \csc \left(\frac{\pi  d}{2}\right) \Gamma \left(\frac{d}{2}-1\right) }{\Gamma \left(3-\frac{d}{2}\right) \Gamma (d-4)} \exp{\!\left[\frac{i}{4}  \left(\pi  d+\frac{4 {x}^2}{t}\right)\right]} \,,
\end{eqnarray}
where 
\begin{align}
    N^d_{O_2}=-16 i \pi^2\left[2^{d-5} (d-2) \pi ^{-d-1} \csc \left(\frac{\pi  d}{2}\right) \Gamma \left(\frac{d}{2}-1\right)^2 \right]^2 .
\end{align}
This expression vanishes for $d=3$, which is understood as the fact that it is a delta function in position space, and is non zero for non integer values of $2<d<4$.

\subsection{$O_3$ in general dimensions}
In this section we present the solution for the $d$-dimensional wave function $\Psi_d(\vec{x}_1, \vec{x}_2, \vec{x}_3)$ for the charge-three operator for $2<d<4$ starting with the free particle Sch\"odinger equation 
\begin{equation}
    -\frac{1}{2} (\sum^3_{i=1}\nabla^2_{\vec{x}_i}) \Psi_d(\vec{x}_1, \vec{x}_2, \vec{x}_3) = E\  \Psi_d(\vec{x}_1, \vec{x}_2, \vec{x}_3).
\end{equation}
As before, the free-particle Hamiltonian can be separated into the centre-of-mass piece and relative coordinates, 
\begin{equation}
    -\frac{1}{2} \Bigl(\frac{1}{3} \nabla^2_{R_{\text{cm}}} + 2(\nabla^2_{r} + \nabla^2_{\rho})\Bigr) \Psi_d(\vec R_{\text{cm}}, \vec{r}, \vec{\rho}) = E \Psi_d(\vec R_{\text{cm}}, \vec{r}, \vec{\rho}), \qquad E= \frac{P_{\text{cm}}^2}{6}+ k^2,
   \end{equation}
where the coordinates $\vec r$ and $\vec \rho$ have been defined as in the main text. We can therefore assume the separation of variables $\Psi_d(\vec{R}_{\rm cm}, \vec{r}, \vec{\rho})= \Psi_d(\vec{R}_{\rm cm})\psi_d^{(3)}(\vec{r}, \vec{\rho})$. The $d$-dimensional Bethe-Pierls boundary condition implies the following differential constraint,
\begin{equation}\label{BP3bodyrrhogend}
    \frac{1}{r^{d-3}}\frac{\partial (r^{d-2} \psi_d^{(3)})}{\partial r}\bigg|_{r=0} =0 .
\end{equation}

We also require that the solution be well behaved as $\rho\rightarrow 0$. As before, we introduce the $d$-dimensional Efimov coordinates,
\begin{equation}
   R = \sqrt{\frac{r^2+\rho^2}{2}}
    ,\quad \alpha=\arctan\frac{r}{\rho}\,,\quad  \hat{r} = \frac{\vec r}{r}\,, \quad \hat{\rho} = \frac{\vec\rho}{\rho}\,,
    \end{equation}
where  $\hat{r}$ and $\hat{\rho}$ are now the $d$-dimensional unit vectors. In these coordinates, we have the hyper radial and hyper angular wave function ansatz as
\begin{equation}\label{3bdywavefngend}
\psi_d^{(3),l_i}(\vec{r}, \vec{\rho})= \frac{F_d(R)}{R^{d-1}}\frac{\phi^{l_i}_{d,s}(\alpha)}{\sin (2\alpha)^{d-2}}Y_{{l_i}}(\hat{\rho}),
\end{equation} 
where we also denote the solution to the $d$-dimensional Laplace-Beltrami operator by the $d$-dimensional spherical harmonics $Y_{{l_i}}$, where $i\in (1,d-1)$. We will be interested in the s-wave answer so for convenience we set $\{l_i\}=0$.

In Efimov coordinates, we obtain the following set of equations for the hyperradial, hyperangular part, 
\begin{subequations}
\begin{align}
     & -F_d''(R)-\frac{1}{R}F_d'(R)+\frac{s^2}{R^2} =2k^2F_d(R) ,\label{hypeqgend2} \\
     & -\frac{2 (d-3) \cos (2 \alpha ) \phi_{d,s}^{0'}(\alpha )}{\sin (2 \alpha )}-(d-3)^2 \phi_{d,s}^0(\alpha )+\phi_{d,s}^{0''}(\alpha )=-s^2 \phi_{d,s}^0(\alpha ).\label{hypeqgend}
\end{align}
\end{subequations}
The choice of ansatz was motivated by the fact that even in general dimensions, the hyperradial equation  continues to be a schrodinger equation in a $\frac{s^2}{R^2}$ potential, where $s$ is the eigenvalue corresponding to the hyperangular equation. To ensure correct symmetry property, we supplement this solution with 
\begin{equation}\label{Psi-d}
 {\Psi_d}^{l_i=0}_{s, k}(\vec{R}_{\rm cm},\vec{\rho},\vec{r})
 =\left(\frac{2}{\sqrt{3}}\right)^{d/2} N_{k,d}^l\Psi_d(\vec{R}_{\rm cm})\frac{F_d(R)}{R^{d-1}}(1-P_{13})\frac{\phi^0_{d,s}(\alpha)}{\sin (2\alpha)^{d-2}} \,,
\end{equation}
where $P_{13}$ implements $\vec{x}_1 \leftrightarrow \vec{x}_3$. The Bethe-Peierls boundary condition \eqref{BP3bodyrrhogend} and the well behavedness in $\rho \rightarrow 0$ limit (i.e $\phi^l_{d,s}(\pi/2)=0$), in these coordinates, translate to a recursion relation condition on the eignevalue $s$.

The hyperradial solution in general dimension therefore is the same as $d=3$, given by \eqref{hyperrad3}. The hyperangular equation in \eqref{hypeqgend} has the following solution which is regular at $\alpha=\frac{\pi}2$ 
\begin{equation}
   \phi^{0}_{d,s}(\alpha)= \cos ^{d-2}(\alpha ) \, _2F_1\left(\frac{1}{2}-\frac{s}{2},\frac{s}{2}+\frac{1}{2};\frac{d}{2};\cos ^2(\alpha )\right).
\end{equation}

Putting this in the generalized Bethe-Peierls condition in eq.~\eqref{BP3bodyrrhogend}, we obtain the following recursion relation for the $\phi_{d,s}$\footnote{More generally, the recursion relation for $s$ can be solved for a system of fermions or bosons ($\eta=-1$, $\eta=2$ respectively)
\begin{equation}
    4\times 3^{d/2} \csc \left(\frac{\pi  d}{2}\right) \cos \left(\frac{\pi  s}{2}\right)+  3\eta\times 2^{d} \, _2F_1\left(\frac{1-s}{2},\frac{s+1}{2};\frac{d}{2};\frac{1}{4}\right)=0.
\end{equation}
For the bosons, this is in agreement with similar recursion relation derived in \cite{Braaten:2004rn} (see eq 411). Our equations match with theirs upon application of the hypergeometric identity 
\begin{equation}
 _2F_1(c-a,c-b;c;z)=(1-z)^{a+b-c} \, _2F_1(a,b;c;z),
\end{equation}
and the identification $s= \sqrt{\lambda}$.}
\begin{equation}\label{s-d}
    4\times 3^{d/2} \csc \left(\frac{\pi  d}{2}\right) \cos \left(\frac{\pi  s}{2}\right)-3\times 2^{d} \, _2F_1\left(\frac{1-s}{2},\frac{s+1}{2};\frac{d}{2};\frac{1}{4}\right)=0.
\end{equation}
In $d=3$, this reduces to the recursion relation obtained in Eq.~\eqref{recrelns3d}.  Equation~(\ref{s-d}) has a solution $s=d-1$ which has to be excluded because it would lead to a zero wave function~(\ref{Psi-d}), which can be seen using the formula $\, _2F_1 (a, b; b; z)=(1-z)^{-a}$. 
We list the nontrival $s$ values for two non integer dimensions in Table \ref{table4}. We have checked that the hyperangular solution is non-zero for these values of $s$ at $\alpha=0$. The normalized wave function takes the form
\begin{equation}
{\Psi_d}_{s,k}^{l=0}(\vec{R}_{\rm cm}, \vec{r}, \vec{\rho})
 = \left(\frac{2}{\sqrt{3}}\right)^{\frac{d}{2}}N^{0}_{k,d} e^{i \vec{P}_{\rm cm}\cdot \vec{R}_{\rm cm}}\frac{J_s(\sqrt{2} k R)}{R^{d-1}} (1-P_{13})\frac{\phi_{d,s}^{0}(\alpha)}{\sin{2\alpha}^{d-2}} \,,
\end{equation}
where
\begin{equation}
N^{0}_{k,d} = \frac{{\sqrt{2} \pi  k }}{R_1 f^0_{s,d} V} \,, \qquad   f^0_{s,d} =  2\int_0\limits^{{\pi}/{2}}\! d\alpha\, \sin^{(d-1)}(2 \alpha) \!\int\! d\Omega_{\hat{\rho}_{d}}\! \int\! d\Omega_{\hat{r}_{d}}  \left|(1-P_{13})\frac{\phi^0_{s,d}(\alpha)}{\sin{2\alpha}^{d-2}}\right|^2 .
\end{equation}
The charge-three operator is similarly defined as 
\begin{align}\label{o3def1gend}
    \langle 0| O^{l_i=0,s'}_{3,d}(\vec{R}_{\rm cm})|\Psi^{(3)}_{l_i=0,s,k,d}\rangle&=\lim_{R\rightarrow 0}  R^{d-1-s'} \!\int\! d\bar{\Omega}_d\,  \Psi_{s',k,d}^{0}(\vec{R}_{\rm cm}, R, \bar{\Omega}_d) (1-P_{13})\frac{\phi_{s,d}^{0}(2\alpha)}{\sin{2\alpha}^{d-2}}\nonumber\\
    &= \left(\frac{2}{\sqrt{3}}\right)^{\frac{d}{2}} N^0_{k,d} f^0_{s,d}\left( \frac{2^{-s/2} k^s}{\Gamma[s+1]}\right)  e^{i \vec{P}_{\rm cm}\cdot \vec{R}_{\rm cm}} .
\end{align}

We obtain the three point function needed for the relevant and irrelevant deformations as 
\begin{eqnarray}\label{spinlo3o2psigend}
      \langle 0| T O^{l_i=0,s}_{3,d}(t,\vec{x}) {O^d}^\dagger_2(t',\yp){\psi_\uparrow^d}^\dagger(t',\xp)|0\rangle &=& \Theta(t-t')N^{l,d}_{3pt} \frac{|\yp-\xp|^{s-1}}{(t-t')^{s+\frac{d}{2}+1}} \exp\!\left[\frac{i3(\vec{x}-\frac{\x'+2\y'}{3})^2}{2(t-t')}+\frac{i(\xp-\yp)^2}{3(t-t')}\right] ,\nonumber\\
      N^{l,d}_{3pt} &=& -\sqrt{2}\left(\frac{2}{\sqrt{3}}\right)^{\frac{d}{2}}\frac{i^{-s-\frac{d}{2}-1} 2^{-\frac{3}{2}-d} 6^{\frac{1-s}{2}} {\phi^*}_{s,d}^{0}(0)}{\pi ^{d/2} \Gamma (s+1)}\,.
\end{eqnarray}

We present the finite piece of the first order irrelevant deformation of charge-three operator two point function in general dimensions $d$, 
\begin{align}
    G^{\text{irrel}}_{ O_{3,d}}(t,\x) =&\,  16i\pi^2 |N^{l,d}_{3pt}|^2 \nonumber\\
    &\times \frac{2^{d/2} \pi ^{d+1} 3^{s-1} \left(d^2-6 d-2 s^2+10\right) \csc \left(\frac{\pi  d}{2}\right) t^{-s-3} \Gamma \left(\frac{d}{2}+s-2\right) }{(d-2) \Gamma \left(3-\frac{d}{2}\right) \Gamma (d-3)}\exp\left[-\frac{i \left(\pi  s t-3 x^2\right)}{2 t}\right].
\end{align}
We see that the finite piece vanishes in $d=3$ because of the $\Gamma(d-3)$ factor in the denominator.

\begin{table}
	\begin{center}
\begin{tabular}{|c|c|c|c|}
			\hline
			$d$ & $s_1$ & $s_2$ & $s_3$ \\
			\hline
			$~2.8~$ & $~2.36961~$ & $~5.09858~$ & $~7.13990~$ \\
			\hline
			3.3 & 1.84220 & 5.15849 & 7.05806\\
			\hline
		\end{tabular}  
	\end{center} 
\caption{The first three values of $s$ in two different fractional dimensions.}
\label{table4}
\end{table}

\section{Irrelevant deformation of the free fixed point}
\label{sec:irrel-free}

In this Appendix, we compute the first order irrelevant deformation about the free fixed point which is described by, 
\begin{equation}
  \mathcal L = \sum_\sigma \psi^\+_\sigma 
   \left( i \d_t + \frac{\nabla^2}2 \right) \psi_\sigma
  - 4\pi a \psi^\+_\uparrow \psi^\+_\downarrow \psi_\downarrow \psi_\uparrow .
\end{equation}

At the free fixed point, the scattering length deformation is an irrelevant deformation. We wish to compute the first-order deformation of the two point function of the charge-three operator. The free field realisation of the charge-three operator describing free neutrons is given by \cite{Hammer_2021}, 
\begin{equation}
    O^{free}_{3i}(t,\x) = \psi_\uparrow \psi_\downarrow \d_i \psi_\uparrow (t,\x) .
\end{equation}

The two-point function of the charge-three operator can be computed by Wick contraction 
\begin{equation}
    \langle T O^{\text{free}}_{3i}(t,\x) O^{\text{free}\dagger}_{3j}(0,\vec{0}) \rangle = \frac{\Theta(t)}{i^{7/2}(2\pi)^{9/2}t^{11/2}} \exp\!\left(\frac{3i x^2}{2 t}\right) \delta_{ij},
\end{equation}
where we have used the free fermion propagator 
\begin{equation}
    \langle T \psi_\sigma(t,\x) \psi^\+_{\sigma'}(0,\vec{0}) \rangle = \frac{\Theta(t)}{(2\pi i t)^{3/2}} {\exp{\left(\frac{i x^2}{2 t}\right)}} \delta_{\sigma\sigma'} .
\end{equation}

The first order irrelevant deformation is similarly computed using Wick contraction, 
\begin{eqnarray}
    \langle T O^{\text{free}}_{3i}(t,\x) O^{\text{free}\dagger}_{3j}(0,\vec{0}) \rangle_a &=& \int d^3\xp \!\!\int\limits_{-\infty}^\infty\!\! dt'\, \langle T O^{\text{free}}_{3i}(t,\x) \left(\psi^\+_\uparrow \psi^\+_\downarrow \psi_\downarrow \psi_\uparrow(t',\xp) \right) O^{\text{free}\dagger}_{3j}(0,\vec{0}) \rangle\nonumber\\
    &=& c^5\Theta(t)\! \int\! d^3\xp\!\! \int\limits_{-\infty}^\infty\!\! dt'\, \Theta(t-t')\Theta(t')\frac{-t'^2 x_i x_j +t ( i t' (t'-t)\delta_{ij}-t x'_i x'_j +2 t' x_i x'_j)}{t^{7/2} t'^4 (t-t')^4}\nonumber\\
    &&\times \exp\!\left[-\frac{2 i \x\cdot\xp}{t-t'}+\frac{i x^2 (3 t-t')}{2 t (t-t')}+\frac{i t x'^2}{t' (t-t')}\right] \nonumber\\
 &=& \Theta(t)\frac{3\delta_{ij} c^5\sqrt[4]{-1} \pi ^{3/2} }{2 t^4 } \exp{\left(\frac{3i x^2}{2 t}\right)} \! \int\limits_{-\infty}^\infty\!\! dt' \, \frac{\Theta(t-t')\Theta(t')}{t'^{3/2} (t-t')^{3/2}}\,,
\end{eqnarray}
where we have used integrals listed in Appendix \ref{usefulint} and $c=(2\pi i)^{-3/2}$. The final integral is divergent and has zero finite piece upon regularisation [using Eq.~\eqref{irrelint1}]. Thus we observe that the first irrelevant correction to the two-point function of the charge-three operator vanishes at the free fixed point.

\bibliography{Arxivv1}

\end{document}